\documentclass[%
 reprint,
 amsmath,amssymb,prx
]{revtex4-1}
\usepackage[bottom]{footmisc}
\usepackage{hyperref}
\usepackage{graphicx}
\usepackage{epstopdf}
\usepackage{dcolumn}
\usepackage{bm}
\usepackage{lipsum}    
\usepackage{amsmath}
\usepackage{mathtools}
\usepackage{color}
\usepackage{soul}

\begin{document}

\title{Evolution of orbital angular momentum in three-dimensional structured light}

\author{Ahmed H. Dorrah$^1$, Carmelo Rosales-Guzm\'an$^2$, Andrew Forbes$^2$, and Mo Mojahedi$^1$}

\affiliation{%
 $^1$ Edward S. Rogers Sr. Department of Electrical and Computer Engineering, University of Toronto, Toronto, ON, M5S 3G4, Canada
}%
 
 \affiliation{%
 	$^2$ School of Physics, University of the Witwatersrand, Private Bag 3, Wits 2050, South Africa
 }
\date{\today}

\begin{abstract}
Light beams with an azimuthal phase dependency of $e^{i\ell\phi}$ have helical phase fronts and thus carry orbital angular momentum (OAM), a strictly conserved quantity with propagation. Here, we engineer quasi three-dimensional (3D) structured light fields and demonstrate unusual scenarios in which OAM can vary locally in both sign and magnitude along the beam's axis, in a controlled manner, under free-space propagation. To reveal the underlying mechanisms of this phenomenon, we perform full modal decomposition and reconstruction of the generated beams to describe the evolution of their intrinsic OAM and topological charge with propagation. We show that topological transition and the associated variation in local OAM rely on the creation, movement, and annihilation of local vortex charges without disturbing the global net charge of the beam, thus conserving the global OAM while varying it locally.  Our results may be perceived as an experimental demonstration of the Hilbert Hotel paradox, while advancing our understanding of topological deformations in general. 
\begin{description}
\item[PACS numbers]
42.25.Bs 42.50.Gy 42.50.Lc 42.65.An
\end{description}
\end{abstract}

\pacs{Valid PACS appear here}
\maketitle

\section{Introduction}
Vortex beams refer to a class of structured light beams \cite{rubinsztein2016roadmap} characterized by azimuthal phase dependency $\sim e^{i\ell\phi}$, where $\ell$ is known as the topological charge of the beam \cite{mandel_wolf_1995,Yao:11,PhysRevA.45.8185}. Such beams possess $\ell$ intertwined helical phase-fronts with an on-axis phase singularity and carry orbital angular momentum (OAM) value of $\ell\hbar$ per photon. The handedness and order of the helical phase twist are determined by the sign and magnitude of $\ell$, respectively. Light's OAM has been utilized in many applications including optical trapping, materials processing, and imaging, and has been extensively reviewed to date \cite{Padgett:17,Padgett2011,Molina-Terriza2007,doi:10.1002/lpor.201100031}. In particular, the unbounded and orthogonal OAM states of light have been extensively deployed in data communications as means of encoding information; both classically and in the quantum regime \cite{Erhard:2018aa,wang2016advances,wang2017data,Ndagano2018,Willner2015}. 

In principle, when optical vortices propagate in a homogeneous isotropic transparent medium, both their spin and orbital angular momenta are conserved  \cite{ref999849741702121,PhysRevLett.96.163905,ALLEN200067,Gregg:15}. In other words, light's OAM is manifested as a strictly conserved quantity, signified by a quantized topological charge $\ell$, and does not change in general under free unperturbed propagation. In very special cases, however, non-trivial topological deformations have been deliberately realized; originally, by interfering vortex modes with Gaussian beams \cite{PhysRevA.56.4064}, then by realizing charge flipping induced in a non-linear medium \cite{Bezryadina:06,Izdebskaya:11}, and in noncanoncial vortices generated by an astigmatic optical setup \cite{Molina-Terriza:01,PhysRevLett.87.023902}. More recently, non-diffracting optical vortices with longitudinally varying topological charge have been observed in air \cite{Davis:16,PhysRevA.93.063864,Yuanjie2018,Otte2018,7787817}, thus opening new opportunities in venues like optical trapping \cite{Ambrosio:15,Ambrosio:16,Pachon:16}, dense data communications \cite{Tian2018}, and remote sensing \cite{Dorrah2018,Dorrah4:18}. In all these developments, a fundamental question on \textit{how OAM conservation is seemingly broken, without violating any laws of physics}, naturally arises. However, a satisfactory answer that resolves such paradoxical behavior, observed in recently developed classes of structured light, has not been provided yet. In addition, the physical dynamics associated with the boundaries at which the light beam undergoes its topological transition has scantly been investigated.

In this work, we examine the evolution of OAM in longitudinally (3D) structured vortex beams --- where the topological charge exhibits unusual, yet controlled, non-trivial transitions with propagation. We start by establishing a general framework for designing and generating such beams, and we consider two case studies in which the sign and magnitude of $\ell$ are changed with propagation. We then perform full modal decomposition and reconstruction of those beams using the Bessel bases. This approach allows us to: a) gain insights into the interplay between the intermodal phases within the beam and its evolution, hence, understanding the underlying mechanism of its topological deformation, b) it allows us to quantitatively measure the OAM density in addition to the local and global values of OAM (and topological charge) of the reconstructed beam as it propagates. As such, we reveal the general mechanism that governs the topological transitions along the beam's axis. We discuss how this mechanism manifests as a practical realization of transfinite mathematics, exhibiting a striking analogy with what is known as the Hilbert's hotel paradox \cite{gamow1988one,potovcek2015quantum,Gbur:16, wang2017hilbert}. We then demonstrate how the OAM, despite its local variation, is always conserved globally --- thus providing a quantitative interpretation addressing the OAM conservation paradox in 3D structured light, which has hitherto been unresolved.

\section{Concept}
Quasi three dimensional structured light, in which the OAM can be longitudinally controlled, is realized here by superimposing multiple vortex modes, $\psi_{\ell}$, of topological charge $\ell$.  Further, each vortex mode itself consists of a superposition of multiple co-propagating beams carrying the same charge $\ell$ but with different spatial frequencies. Without loss of generality, we adopted Bessel vortex beams due to their interesting non-diffracting and self-healing characteristics \cite{doi:10.1080/0010751042000275259}. The  3D structured waveform is expressed as a superposition of $2N+1$ Bessel beams such that
\begin{equation}
\small
\label{Eq1}
\begin{multlined}
U(\rho,\phi,z,t) = \sum_{\ell=-\infty}^{\infty}\psi_{\ell}  \\
= e^{-i\omega t} \sum_{\ell=-\infty}^{\infty}\sum_{m=-N}^{N} A_{\ell, m} J_{\ell}(k_{\rho}^{\ell, m} \rho) e^{i\ell \phi} e^{i k_z^{\ell, m}z}.
\end{multlined}
\end{equation}

Equation~\ref{Eq1} is an exact solution to the paraxial wave equation in which $k_{\rho}^{\ell, m}$ and $k_z^{\ell, m}$ denote the transverse and longitudinal wavenumbers, respectively. The wavenumbers $k_z^{\ell, m}$ are equally spaced in the k-space, in a comb-like setting, with a separation of $2\pi/L$; where $L$ is the desired longitudinal extent of the beam. Additionally, the coefficients $A_{\ell, m}$ represent complex weighting factors for each Bessel beam in the superposition and are calculated by solving \cite{PhysRevA.93.063864}
\begin{equation}
\label{Eq2}
A_{\ell, m} = \frac{1}{L}\int_{0}^{L} F_{\ell}(z) e^{-(i\frac{2\pi}{L}m)z}dz.
\end{equation} 
In the above, the morphological function $F_{\ell}(z)$ defines the desired longitudinal intensity profile of each vortex mode $\psi_{\ell}$ in the superposition of Eq.~(\ref{Eq1}), over the finite longitudinal extent $L$. This is realized by the complex coefficients $A_{\ell, m}$ which represent the projection of $F_{\ell}(z)$ onto the Bessel basis. In essence, the phases and amplitudes of $A_{\ell, m}$ are engineered to shape the longitudinal intensity profile of each mode $\psi_{\ell}$ as it propagates, via controlled interference. 

This method is analogous to Fourier series in which arbitrary periodic waveforms are constructed from a discrete superposition of multiple frequency harmonics weighted by suitable Fourier coefficients. Similarly, here, the discrete superposition of suitably weighted spatial harmonics, $J_{\ell}(k_{\rho}^{\ell, m} \rho) e^{i\ell \phi}$, in Eq.~(\ref{Eq1}) enables us to spatially modulate the envelope of the ensemble via controlled beating among its spatial frequencies. As such, the intensity profile of any desired vortex mode $\psi_{\ell}$ can be designed to \textit{switch on} or \textit{off} along the beam's axis of propagation by virtue of $F_{\ell}(z)$, and the contributions of $\psi_{\ell}$ become spatially variant along the beam's axis. Accordingly, the \textit{effective} charge $\ell$ and intrinsic OAM \cite{PhysRevLett.88.053601} undergo predetermined (and non-trivial) transitions with propagation --- a curious behavior that seemingly breaks OAM conservation in many observation planes. The experimental generation of this class of beam is outlined next.

\section{Methods}
\subsection{Beam Generation}
\begin{figure}[ht]
	\centering
 \includegraphics[width=0.51\textwidth]{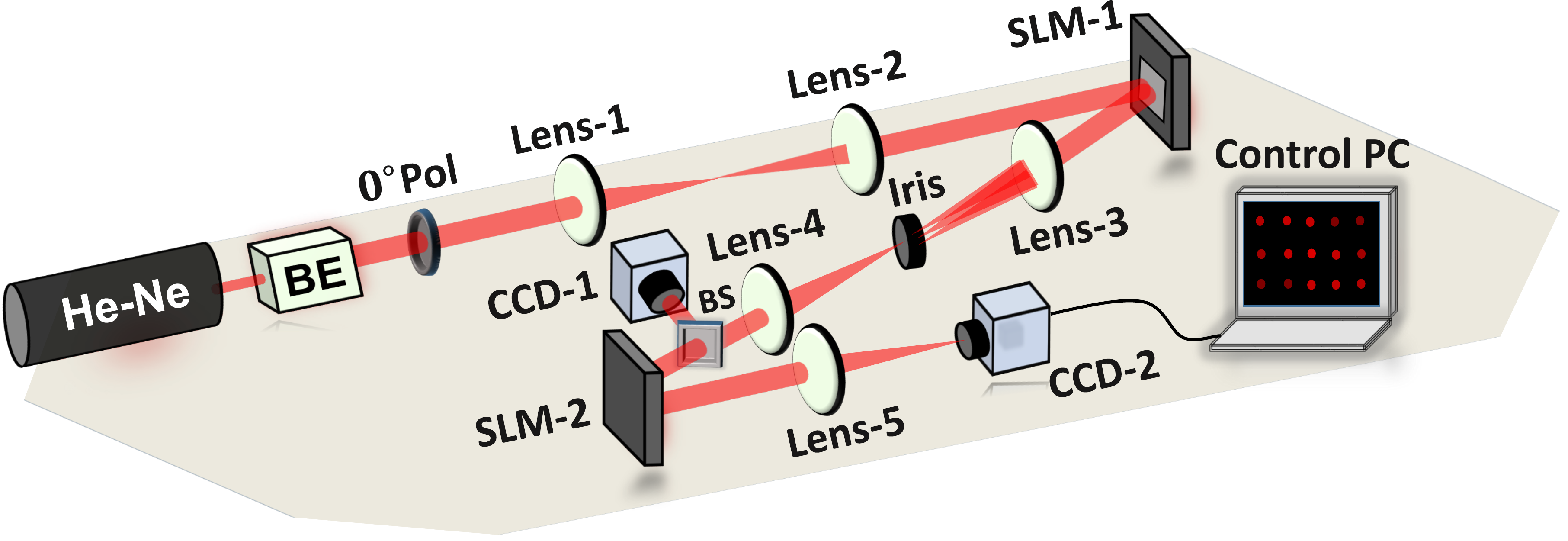}
 
\caption{A holographic setup to create, digitally propagate and then detect 3D structured light. Here, BE refers to the beam expander, Pol is the polarizer, and BS is the beam splitter.}
	\label{Fig1}
\end{figure}

The longitudinally (3D) structured beams were generated using digital holograms realized via programmable spatial lights modulators (SLMs). Figure~\ref{Fig1} illustrates the experimental setup used for beam generation and detection. First, a linearly polarized He-Ne ($\lambda = 632.8$ nm) Gaussian mode was expanded, collimated, and imaged onto SLM-1 (LCOS reflective phase SLM with 1920$\times$1080 resolution and 8 $\mu$m pixel pitch). Next, following the modulation scheme described in Refs. \cite{Arrizon:07,rosales2017shape}, the desired pattern given by Eq.~(\ref{Eq1}) was encoded into a 2D hologram that was displayed on SLM-1. In addition, a 2D grating function was encoded on the hologram to spatially separate the generated signal in the $k$-space from the on-axis carrier. As such, the reflected beam from SLM-1 was imaged and spatially filtered using a $4f$ system incorporating an iris to remove unwanted diffraction orders. A CCD camera (CCD-1) was then used to monitor the transverse intensity profile of the resulting beam with propagation. Finally the detection module, comprised of SLM-2 and a 2-f system, was used to decompose the propagating beam as discussed next.

\subsection{Modal Decomposition and Reconstruction}
To examine the physical dynamics associated with the spatial evolution of OAM, we performed full modal decomposition of $U(\rho,\phi,z,t)$ along its propagation direction. Recall that any optical field can be expressed as a coherent superposition of linearly independent basis functions $V_j$ such that 
\begin{equation}
\tilde{U}({\rho,\phi})=\sum_{j=1}^P c_j V_j({\rho,\phi}),
\label{Eq3}
\end{equation}
where $c_j=|c_j|e^{\Delta\theta_j}$, whereas $\Delta\theta_j$ is the intermodal phase between $V_j$ and a reference mode $V_0$ such that $\Delta\theta_j=\theta_j-\theta_0$. Further, here, the field is decomposed at a fixed $z$-plane (i.e. $z_k$) such that $U(\rho,\phi,z=z_k,t) = \tilde{U}({\rho,\phi})e^{-i\omega t}$. The modal power coefficients, $|c_j|^2$, are mathematically obtained by evaluating the inner product between $\tilde{U}$ and each basis $V_j$, i.e. $|c_j|^2 = |\langle \tilde{U}|V_j\rangle|^2$, given by 
\begin{equation}
|c_j|^2 = \left| \int\int_{\mathbb{R}^2} V_j^*({\rho,\phi})\cdot \tilde{U}({\rho,\phi}) \rho d\rho d\phi \right|^2,
\label{Eq4}
\end{equation}
\noindent and normalised such that $\sum_j |c_j|^2 = 1$. Importantly, the inner product given by Eq.~(\ref{Eq4}) can be performed all-optically by projecting $\tilde{U}(\rho,\phi)$ onto correlation filters digitally encoded on a programmable SLM --- a technique that is also compatible with our setup \cite{Kaiser:09,Flamm:12,Dudley:14,1367-2630-15-7-073025,Flamm2013-2,rosales2017shape}. In this case, SLM-2 acts as a digital filter that performs an inner product between $\tilde{U}(\rho,\phi)$ and the basis function $V_j$. This requires the complex conjugate of $V_j$ to be encoded on the display of SLM-2, as follows 
\begin{equation}
H_j({\rho,\phi})=V_j^*({\rho,\phi}).
\end{equation}
In practice, the outcome of this inner product is obtained by performing an on-axis intensity measurement in the far field of the reflected beam --- realized here by placing Lens-5 in a $2f$ configuration thus transforming the plane of SLM-2 into the far field (k-space) to be detected by CCD-2. As such, the modal power coefficients $|c_j|^2$ can be obtained. Note that the absolute weights $|c_j|$, solely, are not sufficient to fully reconstruct the pattern $\tilde{U}(\rho,\phi)$. In fact, the intermodal phases $\Delta\theta_j$ are also needed, and are readily obtained from two interferometric measurements that are feasibly enabled using SLM-2 as well \cite{Kaiser:09,Flamm:12,Dudley:14,1367-2630-15-7-073025,Flamm2013-2,rosales2017shape}. This requires two additional correlation functions to be encoded on SLM-2; namely, $H_j^{\mbox{cos}}=[V_j^*+V^*_0]/\sqrt{2}$ and $H_j^{\mbox{sin}}=[V_j^*+iV^*_0]/\sqrt{2}$. Accordingly, the intermodal phases $\Delta\theta_j$ are evaluated from 
\begin{equation}
\small
\Delta\theta_j=-\mathrm{arctan}\left[\frac{2I_j^{\mathrm{sin}}-|c_j|^2 - |c_0|^2}{2I_j^{\mathrm{cos}} - |c_j|^2 - |c_0|^2}\right].
\label{phase}
\end{equation}

\noindent Here, $I_j^{\mbox{sin}}$ and $I_j^{\mbox{cos}}$ depict the measured on-axis intensity signals resulting from the inner products with $H_j^{\mbox{sin}}$ and $H_j^{\mbox{cos}}$, respectively \cite{Kaiser:09,Flamm:12,Dudley:14,1367-2630-15-7-073025,Flamm2013-2,rosales2017shape}.

A key benefit in using SLMs as digital filters is the ability to combine multiple correlation filters, encoded with different grating periods, and to multiplex them into a single hologram. This allows to spatially separate the respective inner products, in the Fourier plane, thus reducing the total number of required measurements. By adding all the elements of the bases functions $V_j$ weighted by their respective complex coefficients $|c_j|e^{\Delta\theta_j}$, as per Eq.~(\ref{Eq3}), the transverse pattern at $z=z_k$, (i.e. $\tilde{U}(\rho,\phi)$), can be reconstructed. This process is then repeated at different $z$-planes to fully reconstruct the 3D structured field, $U(\rho,\phi,z,t)$.

We note that precise alignment between SLM-1 and SLM-2 is particularly critical when performing the optical correlations (inner products). For this purpose, both SLMs were mounted on 3D tranlational stages with micron-scale resolution and were fixed throughout the measurements. As such, beam propagation has been realized digitally by updating the distance $z$, in the propagation term of Eq.~(\ref{Eq1}), with increments of 1 cm. This approach provides an accurate realization of beam propagation within the interval $z\leq L$, under the paraxial regime. It also ensures consistency in the alignment when recording the successive planes, as opposed to mechanically displacing the detection system. 

\subsection{Choice of Bessel functions}
In principle, one may represent the optical field $U(\rho,\phi,z,t)$ in terms of a given set of basis functions. In our case, at first glance, the Bessel functions may seem a natural choice for basis functions, given that $U(\rho,\phi,z,t)$ itself is constructed from a discrete superposition of Bessel modes. However, this choice is accompanied with an inherent challenge: a necessary condition for the all-optical inner product discussed previously is to adopt orthonormal basis functions \cite{Flamm2013-2} --- a requirement that is not generally satisfied for Bessel functions with different spatial frequencies $k_{\rho}$. Orthogonality implies minimal overlap between the modes such that
\begin{equation}
\int\int_{\mathbb{R}^2} V^*_{j}\cdot V_{j'} \rho d\rho d\phi=\delta_{j,j'},
\end{equation}
whereas normalization ensures that the total energy of the input field is conserved when transformed from one function space to another. In this case $\sum|c_j|^2=1$. Violating the orthonormality condition can lead to inaccurate reconstruction of the field and may violate the conservation of energy in some cases \cite{Flamm2013-2}. To satisfy the orthogonality requirement we first studied the overlap between the Bessel functions with different values of $k_{\rho}$. In this regard, we performed multiple cross-talk measurements among the Bessel modes at different separations between their wavenumbers ($k_\rho$) \cite{Trichili:14}. More details on the cross-talk analysis can be found in Appendix A. Based on this analysis, we were able to set a limit on the minimum separation between the values of $k_\rho$ so that the generated beam $U(\rho,\phi,z,t)$ contains a discrete set of \textit{pseudo-orthogonal} Bessel functions --- which will then be encoded in the correlation filters at the detection stage. In essence, prior knowledge of $U(\rho,\phi,z,t)$ allows us to narrow down the number of digital filters encoded on SLM-2, thus making the modal decomposition process more efficient. Additionally, the Bessel bases are normalized with respect to their total energies (which are mode dependent) to satisfy the normality condition. 
 
A powerful advantage in modal decomposition is that it provides full access to key physical quantities of the reconstructed beam, such as its Poynting vector, OAM density, and effective charge $\ell$. Furthermore, the intermodal phases provide detailed insights into the mechanisms governing the 3D structured beam evolution and topological deformation. In the following section, we showcase two scenarios in which the sign and magnitude of $\ell$ vary with propagation. We then describe the evolution of the aforementioned physical quantities with propagation and show how the OAM conservation is always satisfied.

\section{Results}
We have experimentally generated two different light patterns constructed from a superposition of the vortex modes $\psi_{\ell}$. Each mode $\psi_{\ell}$ is composed of 5 Bessel beams (N=2). The longitudinal wavenumbers $k_{z}^{\ell, m}$ were centered around $0.999995 k_0$ ($k_0=2\pi/\lambda_0$), and were equally spaced with separations of $4\pi$ in the k-space. This choice of $k_{z}^{\ell, m}$ yield Bessel beams whose transverse wavenumbers $k_{\rho}^{\ell, m}$ are sufficiently separated to ensure low cross-talk among the Bessel modes ($<$-10 dB). This becomes particularly useful for fulfilling the orthogonality requirement in the process of modal decomposition and reconstruction as discussed in the previous section (additional details can be found in Appendix B). To this end, we generated and fully reconstructed two structured beams in which the sign and magnitude of the topological charge were made to vary with propagation. We start by discussing the first pattern as described below.
\subsection{Pattern 1: Reversing the sign of the topological charge}

In the first experimental scenario, the propagating field $U(\rho,\phi,z,t)$ flips its topological charge from $\ell=2$ to $\ell=-2$ as it propagates. In this case, $U(\rho,\phi,z,t) = \psi_2 + \psi_{-2}$. The morphological function $F_{\ell}(z)$ associated with each vortex mode $\psi_{\ell}$ was defined as  
\begin{equation}
\label{Eq8}
\begin{split}
\small
F_{\ell}(z) \begin{cases}
F_{2}  = 1 \ \ \ \ \ \ \ \ \ \ \ \ 0 \ $cm$\ \leq z \leq 14 \ $cm$,    \\
F_{-2} = 1 \ \ \ \ \ \ \ \ \ 14 \ $cm$\ \leq z \leq 28 \ $cm$,    \\
F_{2}  = F_{-2} = 0 \ \ \ \ \ \ \ \ \ \ \ \ \      $elsewhere$.
\end{cases}  
\end{split}
\end{equation} 

\begin{figure}[ht!]
	\centering
	\includegraphics[width=0.45\textwidth]{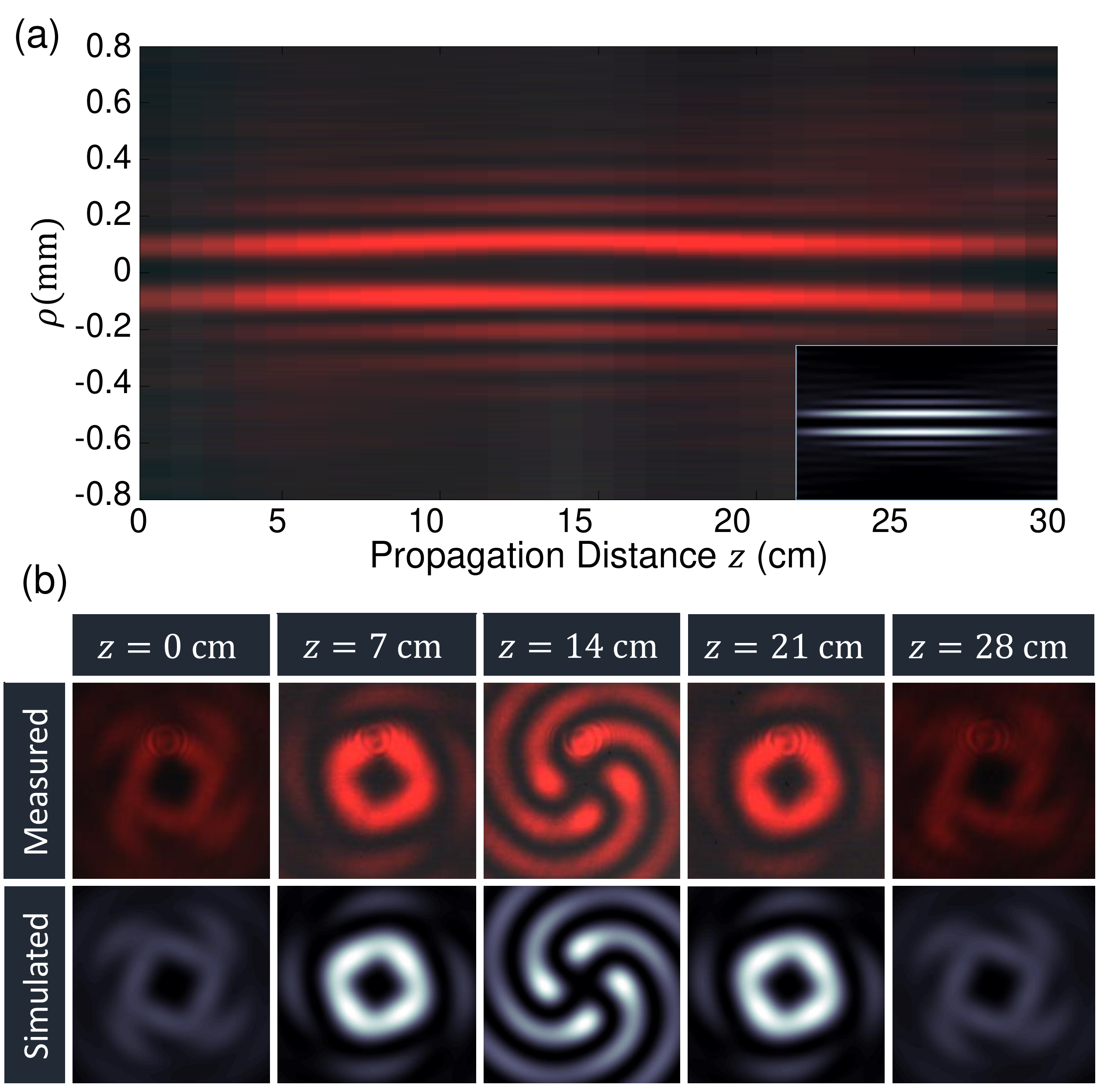}
   
\caption{Measured and simulated intensity of the beam reversing its charge. a) Measured longitudinal intensity profile (the inset depicts the simulated intensity profile). b) Measured and simulated transverse intensity profiles obtained at propagation distances: $z=0$ cm, $z=7$ cm, $z=14$ cm, $z=21$ cm, and $z=28$ cm.}
	\label{Fig2}
\end{figure}

According to Eq.~(\ref{Eq8}), the beam is designed to possess a topological charge $\ell=2$ over the region (0 cm  $\leq z \leq$ 14 cm), and then undergoes topological inversion at $z=14$ cm, at which the charge is reversed to become $\ell=-2$ for the remaining distance (14 cm  $\leq z \leq$ 28 cm). The complex amplitude hologram for this beam has been computed based on Eq.~(\ref{Eq1}) and the procedure described in Ref.~\cite{Arrizon:07}, then displayed on SLM-1. The corresponding intensity profiles, measured by CCD-1, are depicted in Fig.~\ref{Fig2}. The longitudinal intensity profile in Fig.~\ref{Fig2}(a) has been rendered by aggregating 1D slices obtained from the transverse CCD images taken over 30 cm (with 1 cm steps) in the $z$ direction. The rendered profile is in good agreement with the simulated results shown in the inset. The transverse profiles are displayed in Fig.~\ref{Fig2}(b) at various propagation distances along the $z$ direction. Note that there exist a region in space, at $z=14$ cm, where the vortex mode collapse from a closed ring into a petal-like structure. This breakdown is a signature of topological inversion as will be shown later.  

\begin{figure}[ht!]
	\centering
	\includegraphics[width=0.49\textwidth]{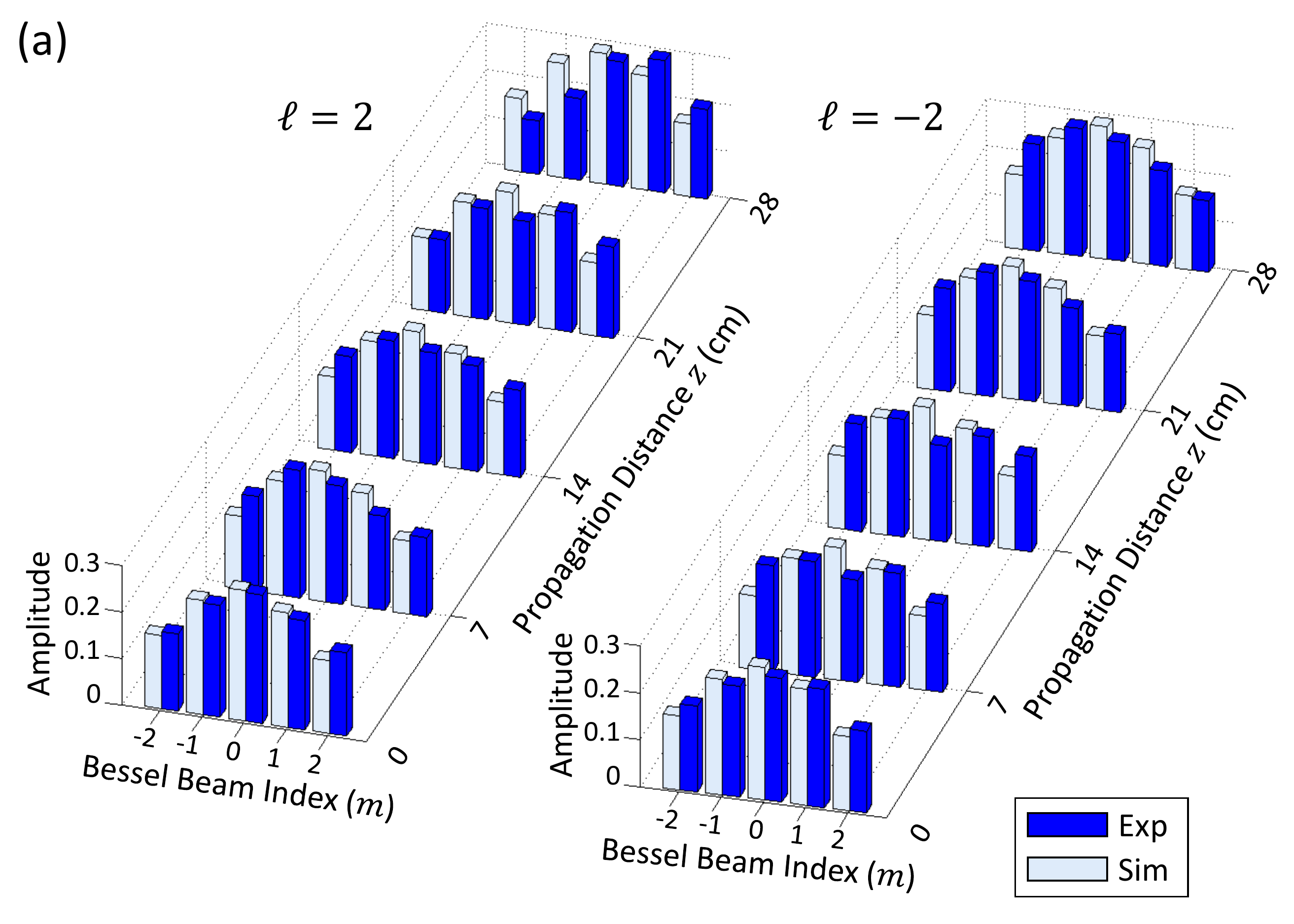}
	\includegraphics[width=0.49\textwidth]{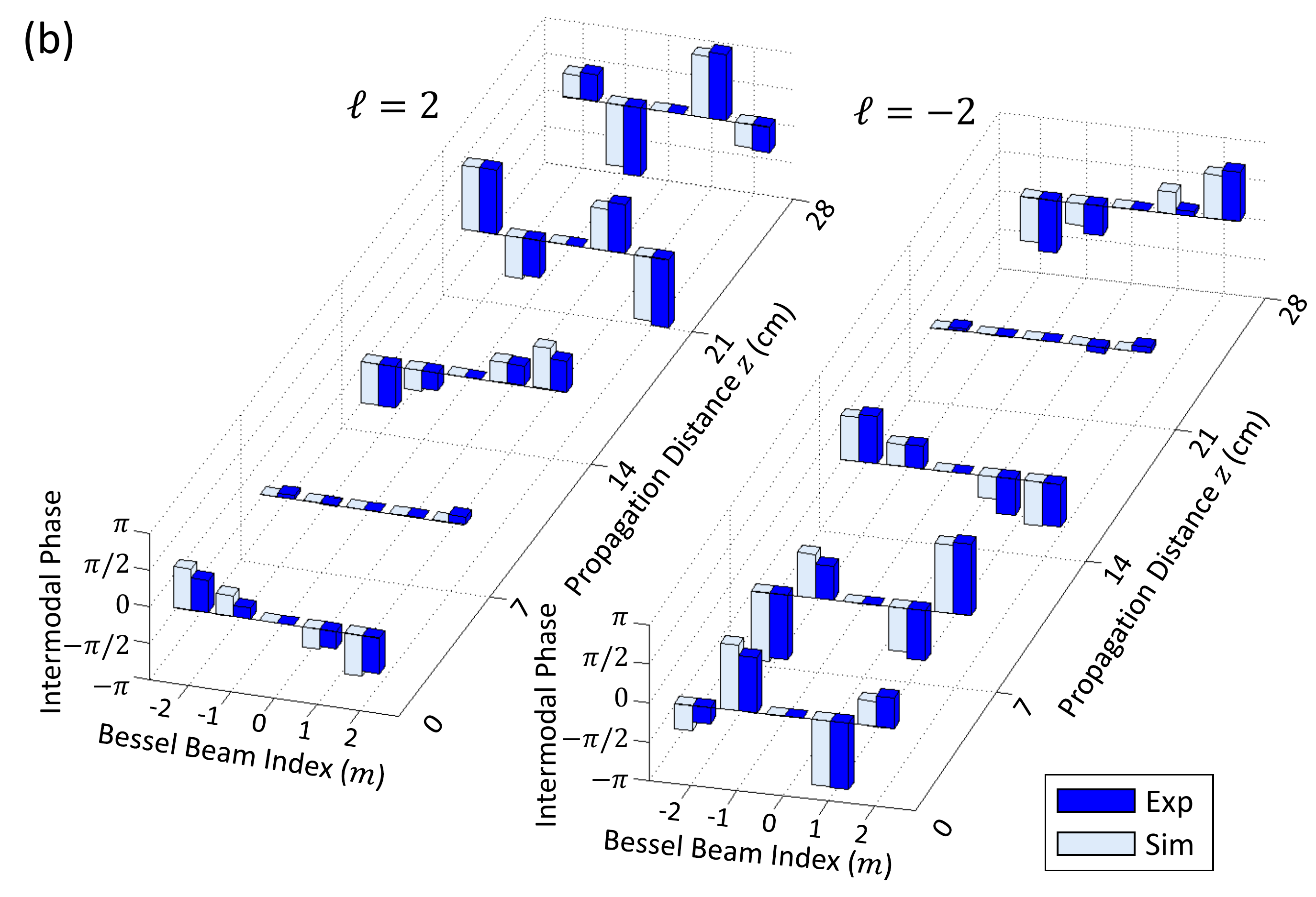}

\caption{Spectral decomposition of the generated 3D structured light into Bessel basis. a) Amplitudes of $A_{2, m}$, and $A_{-2, m}$. b) Intermodal phases among the Bessel modes within $\psi_2$ and $\psi_{-2}$. The coefficients are displayed at $z=0$ cm, $z=7$ cm, $z=14$ cm, $z=21$ cm, and $z=28$ cm.}
	\label{Fig3}
\end{figure}

The intensity measurements alone, however, are not sufficient to examine key properties of the generated beam such as its OAM and topological charge. To access these quantities, we performed full modal decomposition and reconstruction of $U(\rho,\phi,z,t)$ into its Bessel basis. The measured amplitudes of the Bessel beams correspond to the absolute values $|A_{\ell, m}|$, whereas the measured intermodal phases accounts for both the phases associated with $A_{\ell, m}$ in addition to the phases accumulated with propagation. More specifically, the intermodal phases describe the differences among the phases of $A_{\ell, m}e^{i k_z^{\ell, m}z}$. Note that phase accumulation is dependent on the spatial frequency of each Bessel beam in the superposition. 

The measured amplitudes and intermodal phases among the Bessel beams are shown in Fig.~\ref{Fig3} in comparison with the theoretical predictions. Here, we show the coefficients at five distances: $z=0$ cm, $z=7$ cm, $z=14$ cm, $z=21$ cm, and $z=28$ cm, chosen as representative samples for the beam evolution. In principle, the amplitudes of $A_{\ell, m}$ remain constant throughout the beam's propagation (as long as it is not perturbed), as seen from Fig.~\ref{Fig3}(a). On the other hand, the intermodal phases vary with propagation and play a critical role in shaping the longitudinal beam profile. For instance, at $z=0$ cm, the Bessel beams within $\psi_{-2}$ are out-of-phase and, hence, destructively interfere and do not contribute to the beam's intensity at the center. As the beam propagates, the intermodal phases evolve such that, at $z=7$ cm, the Bessel beams within $\psi_2$ become in-phase and, hence, interfere constructively in the beam's center while, at the same $z$ position, the Bessel beams within $\psi_{-2}$ are still out-of-phase and destructively interfere. 

Destructive interference in the vortex mode $\psi_{-2}$ implies that its energy is dispersed in the outer rings of the beam. In this case, the topological charge in the beam's center is predominately $\ell=2$ as a result of the contributions of $\psi_2$, as will be verified shortly. This state of intermodal phases is reversed later on. For example, at $z=21$ cm, the Bessel beams associated with $\psi_{-2}$ become in-phase whereas those associated with $\psi_2$ become out-of-phase. Hence, the beam's topology is reversed. Evidently, there exist a transitional region midway along the beam's axis (at $z=14$ cm) where neither contributions of $\psi_2$ nor $\psi_{-2}$ vanish. At this position, there is an overlap between $\psi_2$ and $\psi_{-2}$ contributions and topological inversion takes place. Finally, at $z=28$ cm, the Bessel beams within $\psi_2$ and $\psi_{-2}$ become out-of-phase and the 3D structured beam gradually switches off.

\subsection{Mechanism of Topological Charge Inversion}

\begin{figure*}
	\centering
	\includegraphics[width=1\textwidth]{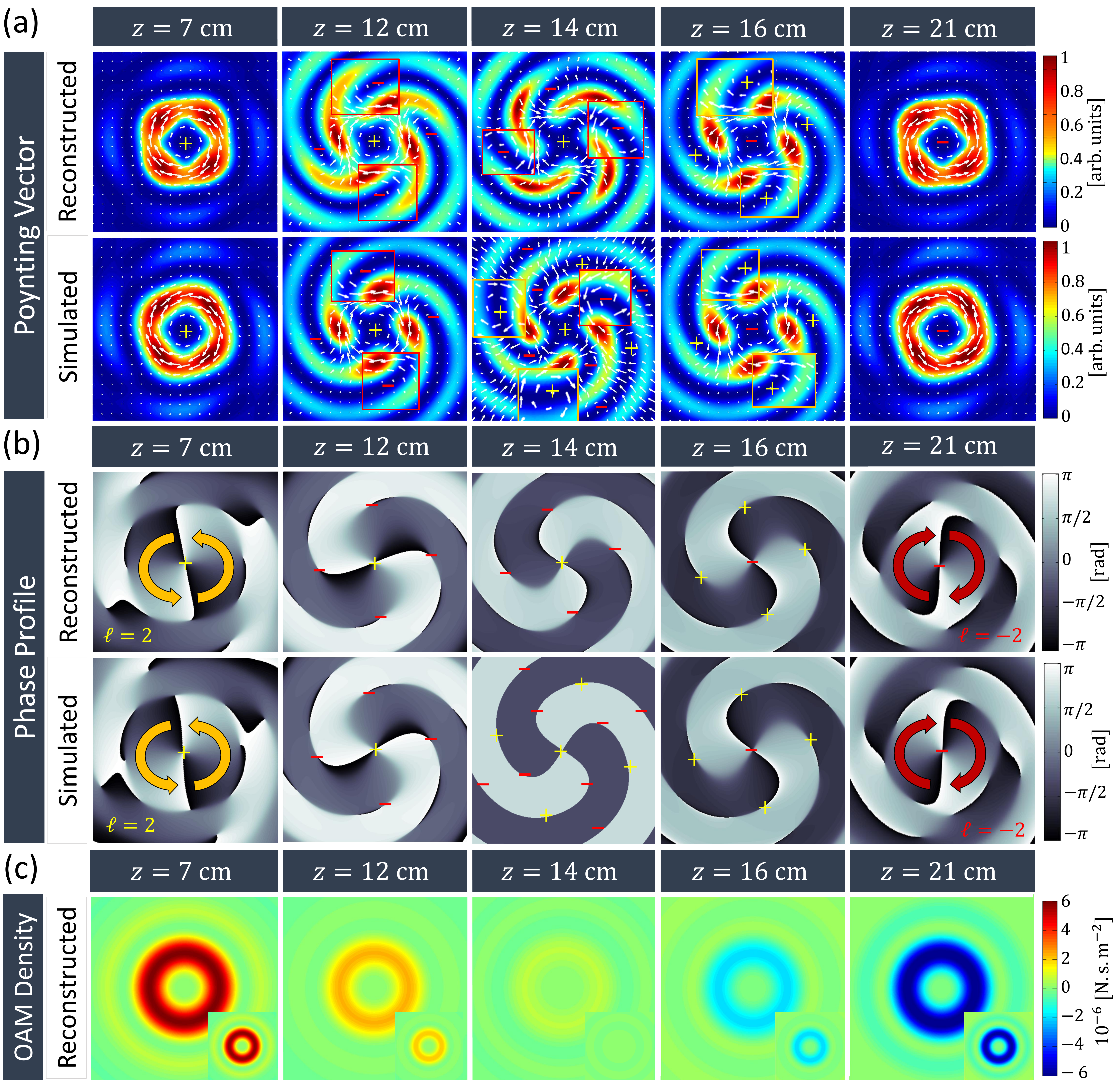}
    \caption{Theoretical and experimental evolution of the Poynting vector, phase, and OAM density of our 3D structured light reversing its charge.  All plots reveal the transverse plane at various propagation distances, $z$.  a) Transverse components of the Poynting vector. The white arrows trace the direction of energy flow and the `+' and `-' signs denote the polarity of the local vortices. The boxed regions represent enlarged sections of the image. b) Reconstructed and simulated phase profile. The arrows depict the sense of helicity in the phase and their number denote the topological charge. c) OAM density profiles reconstructed at different propagation distances. The insets depict the simulated profiles. The OAM density evolves from positive to negative distribution as the beam propagates.}
  	\label{Fig4}
\end{figure*}

To investigate the topological inversion more closely we compare the measured Poynting vector (inferred from modal reconstruction) to the theoretically predicted one. Recall that the Poynting vector points in the direction perpendicular to the wavefront \cite{PhysRevA.45.8185} and is given by the following expression for scalar beams \cite{Litvin:11}
\begin{equation}
\mathbf{S}=\frac{\epsilon_0\omega}{4}\left[ i(U\nabla U^*-U^*\nabla U)+2k_0|U|^2\hat{\mathbf{z}}\right].
\end{equation}
Here, $\epsilon_0$ is the free space permittivity and $\omega$ is the angular frequency. In vortex modes, the Poynting vector typically follows a spiraling path with propagation (perpendicular to the rotating helical wavefront) and, hence, it has non-zero transverse components. Figure~\ref{Fig4}(a) shows the transverse components of the Poynting vector of the reconstructed beam at various planes along its propagation direction, and compared with the simulated results. At $z=7$ cm, it is observed that the energy of the vortex beam predominately flows in the clock-wise direction. This behavior is then disrupted at $z=14$ cm, after which the energy reverses its flow to the counter clock-wise direction (as seen at $z=21$ cm). Topological inversion is confirmed by looking at the reconstructed phase profiles depicted in Fig.~\ref{Fig4}(b), from which it is evident that the phase inverts its helicity from $\ell=2$ to $\ell=-2$ as the beam propagates. 

Reversal of the topological charge hinges on several intriguing dynamics. First, the vortex beam collapses from a closed ring (with charge $\ell=2$) to a petal-like structure. This transformation is a consequence of overlapping vortex modes with opposite helicities ($\psi_2$ and $\psi_{-2}$) over the same space region. Second, the phase-front gradually loses its helicity until it becomes unfolded (flat) at $z=14$ cm. This \textit{unwrapping} of the phase-front is associated with the formation of chains of smaller vortices with alternating signs located along the binary phase dislocations, as marked by `+' and `-' signs. Note that chains of vortices with alternating signs are typically a signature of fractional (non-integer) vortex beams \cite{1464-4258-6-2-018,1367-2630-6-1-071,2040-8986-20-4-045607}, and are experienced here during the topological inversion. Third, the vortices with `-' sign approach the beam center, guided along the path of phase dislocation, and coalesce into one large vortex that replaces the $\ell=2$ vortex, thus reversing its topology. Simultaneously, the $\ell=2$ vortex is divided into smaller vortices that exit from the beam center towards its outer ring. Next, the phase-front gradually acquires a helical nature but in the opposite sense ($\ell=-2$). Finally, after all the charges have been judiciously transported to and from the beam's center (guided on the path of phase dislocations), the `+' charges annihilate and the petal-like structure closes again into a ring, this time with charge $\ell=-2$.

The orbital angular momentum density in the $z$-direction for this beam is obtained from \cite{Litvin:11}
\begin{equation}
\mathit{j_z}=(\rho\hat{\mathbf{\rho}}\times\frac{\mathbf{S}}{c^2})_z.
\label{Eq10}
\end{equation}
Figure~\ref{Fig4}(c) illustrates the distribution of OAM density in the beam center as the beam propagates. The OAM density evolves from positive values at $z=7$ cm (associated with $\ell=2$), to acquire negative values with same magnitude and opposite sign at $z=21$ cm --- as expected from a vortex beam with $\ell=-2$. Interestingly, there exist an overlap region midway, at $z=14$ cm, where the OAM density is dispersed over concentric rings with alternating signs in agreement with Poynting vector and phase pictures discussed above.

\subsection{Conservation of OAM and charge}
\begin{figure}[ht!]
	\centering
	\includegraphics[width=0.5\textwidth]{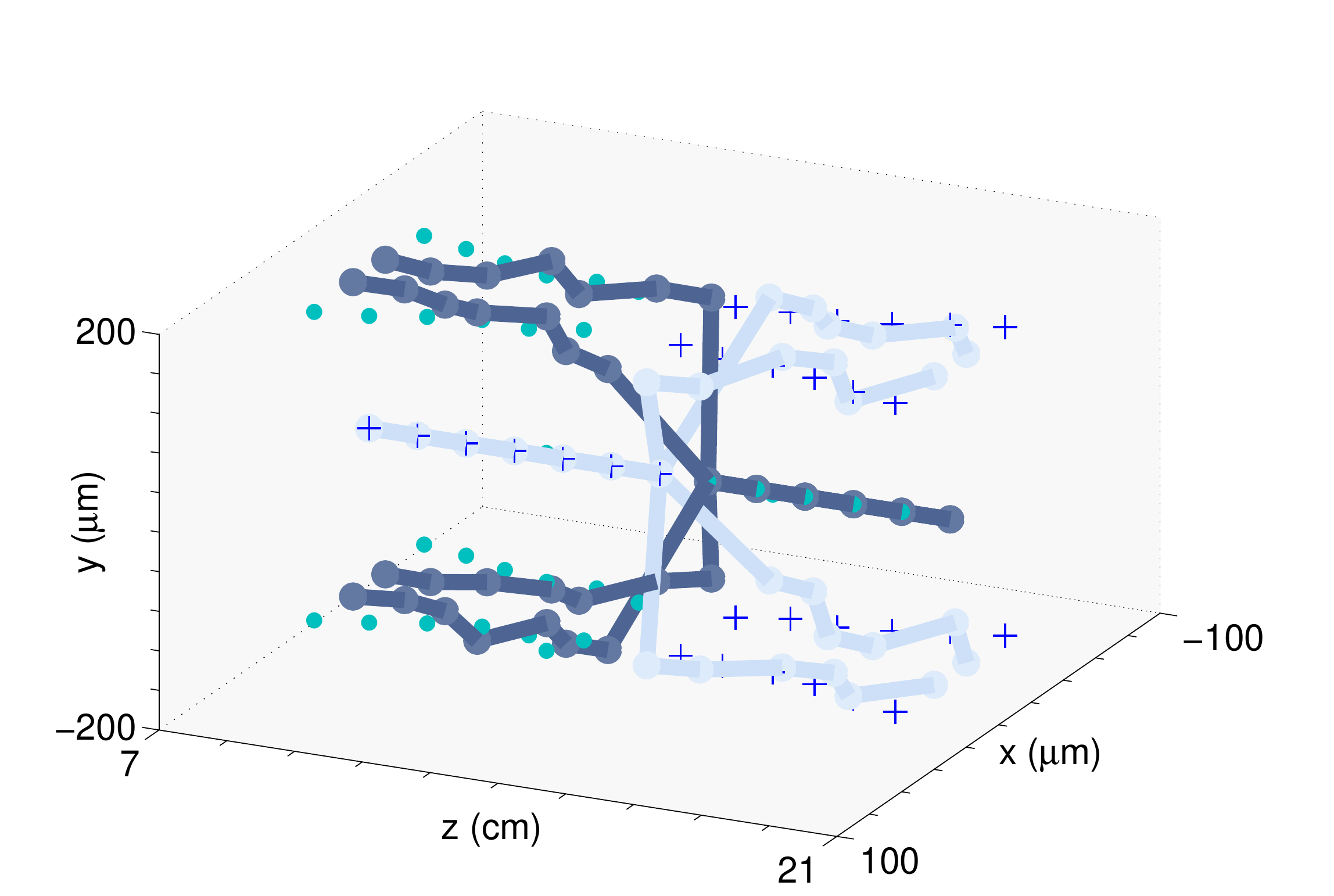}
\caption{Movement of charges in the vicinity of the beam's center. The solid lines depict the experimentally reconstructed paths whereas the markers represent the theoretical paths.}
	\label{Fig5}
\end{figure}

Thus far we have described the mechanism governing the topological inversion from $\ell=2$ to $\ell=-2$ in our 3D structured beam. The transition occurs as a result of judicious creation, movement, and annihilation of phase singularities as the beam propagates. The evolution of the topological charge is further illustrated in Fig.~\ref{Fig5} which depicts the 3D path traversed by the charges in the vicinity of the beam's center as it undergoes the topological transition. The reconstructed paths have been obtained by tracking the location of phase singularities near the beam's center. 

The plot emphasizes the role of the outer rings, acting as a reservoir that injects new charges into the beam's center, when needed, to satisfy the desired OAM profile dictated by Eq.~\ref{Eq8}. Notice how the negative charges (in turquoise markers) are injected into the beam's center while the positive charges exit from the center into the outer rings of the beam to realize the topological inversion at $z=14$ cm.  

A fundamental question on how the OAM and charge are conserved, in such case, naturally arises. To provide a quantitative answer, we computed the OAM and effective charge under two aperture sizes. The total orbital angular momentum $J_z$ per unit length is evaluated by integrating the OAM density over the transverse plane of the beam according to
\begin{equation}
J_z=\int\int_\mathbb{R}\mathit{{j_z}}\hat{\mathbf{z}}\cdot d\mathbf{A}.
\label{Eq11}
\end{equation}

In addition, the effective topological charge of the beam is inferred from the ratio between its OAM (${J_z}$) and Energy per unit length ($W$) such that \cite{doi:10.1002/9783527635368.ch1}
\begin{equation}
\frac{{J_z}}{W}=\frac{\int\int_\mathbb{R}(\mathbf{\rho}\times \langle\mathbf{E}\times \mathbf{B}\rangle_z)\cdot d\mathbf{A}}{c\int\int_\mathbb{R}\langle\mathbf{E}\times \mathbf{B}\rangle_z\cdot d\mathbf{A}} = \frac{\ell_{\mbox{eff}}}{\omega}.
\label{Eq12}
\end{equation}

\begin{figure}[ht!]
	\centering
	\includegraphics[width=0.42\textwidth]{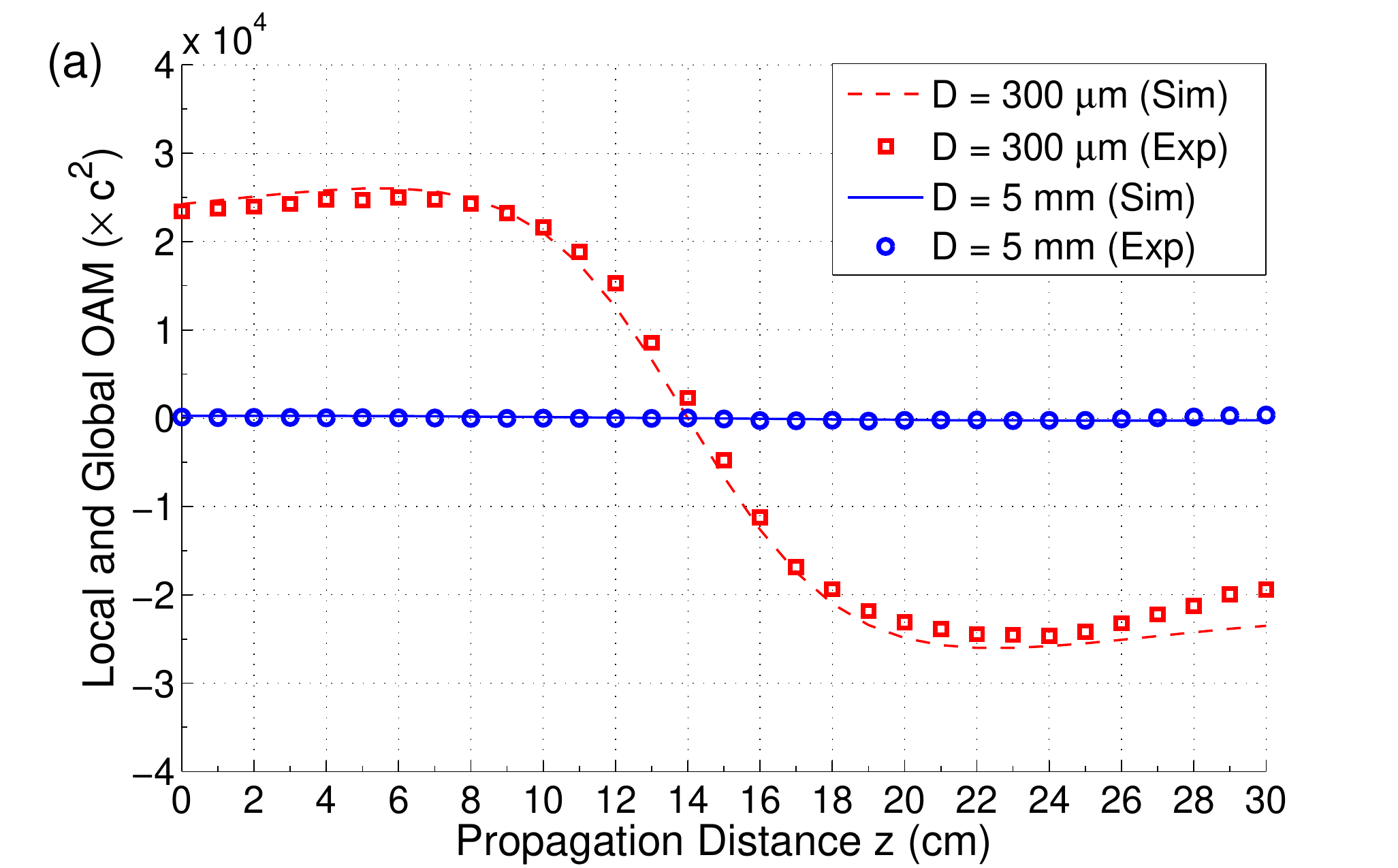}
	\includegraphics[width=0.42\textwidth]{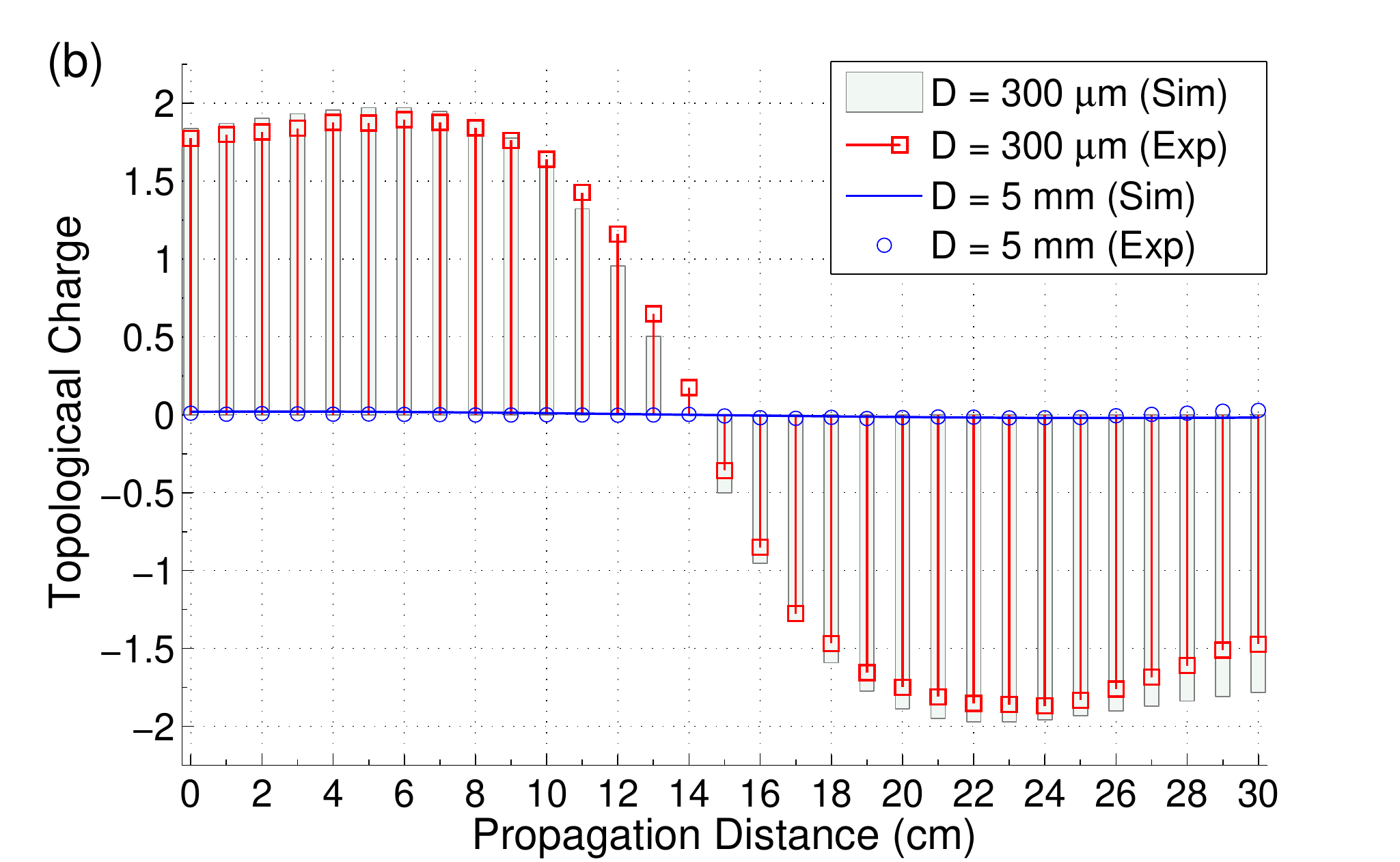}
  
\caption{Evolution of the OAM and topological charge considering two aperture diameters $D=300$ $\mu$m, and $D=5$ mm. a) Local and global OAM of the beam obtained along its direction of propagation. b) Local and global effective topological charge evaluated along the beam's axis.}
	\label{Fig6}
\end{figure}

The OAM per unit length ($J_z$) and effective topological charge are depicted in Fig~\ref{Fig6}(a) and (b), respectively. The experimental results are compared with the simulations and are in very good agreement. We note that these quantities have been obtained over two different aperture diameters:  a) $D=300$ $\mu$m, and b) $D=5$ mm. When computed over the smaller aperture, the OAM exhibits a transition from positive to negative quantities along the propagation direction, and it reaches an inflection point characterized by zero-OAM in between. Evidently, the OAM conservation is seemingly broken over this finite section of the beam. We refer to this quantity as the \textit{local} OAM of the beam. Interestingly, when the OAM is evaluated across the entire cross section of the beam ($D=5$ mm) it maintains the same value regardless of the propagation distance $z$. The \textit{global} OAM is always zero in this case. These observations quantitatively establish that while the OAM may vary locally, the global OAM is always conserved. 

A similar picture can be seen when evaluating the effective topological charge $\ell_{\mbox{eff}}$ under two aperture sizes, as shown in Figure~\ref{Fig6}(b). It is noticed that the local charge evolves from $\ell_{\mbox{eff}}=2$ to $\ell_{\mbox{eff}}=-2$ when observed over the smaller aperture while the global $\ell_{\mbox{eff}}$ remains zero at all planes along the $z$ direction. Therefore, topological inversion occurs without violating neither the OAM nor the charge conservation. While the number of charges confined in a sub volume of the beam can vary, the net charge throughout the entire volume remains in balance. In essence, charges are created in pairs across the beam and topological inversion leverages on the deliberate movement of some of those charges towards the beam's center leaving the opposite charges in the outer rings of the beam.    

To compliment these findings, in the next section, we present another scenario in which the magnitude of the topological charge can be made to vary with propagation.

\subsection{Pattern 2: Changing the magnitude of the topological charge}

In the second case study, the propagating field $U(\rho,\phi,z,t)$ was designed to change the magnitude of its topological charge from $\ell=1$ to $\ell=3$ as it propagates. For this scenario, the propagating waveform is expressed as $U(\rho,\phi,z,t) = \psi_1 + \psi_{3}$. The morphological function $F_{\ell}(z)$ associated with each vortex mode $\psi_{\ell}$ is chosen as 
\begin{equation}
\begin{split}
\small
F_{\ell}(z) \begin{cases}
F_{1}  = 1 \ \ \ \ \ \ \ \ \ \ \ \ \ \ 0 \ $cm$\leq z \leq 11 \ $cm$,    \\
F_{3} = 1.15 \ \ \ \ \ \ \ \ \ \ 11 $cm$\leq z \leq 23 \ $cm$,    \\
F_{1}  = F_{3} = 0 \ \ \ \ \ \ \ \ \ \ \ \ \      $elsewhere$.
\end{cases}  
\end{split}
\label{Eq13}
\end{equation} 

\begin{figure}[ht!]
	\centering
	\includegraphics[width=0.45\textwidth]{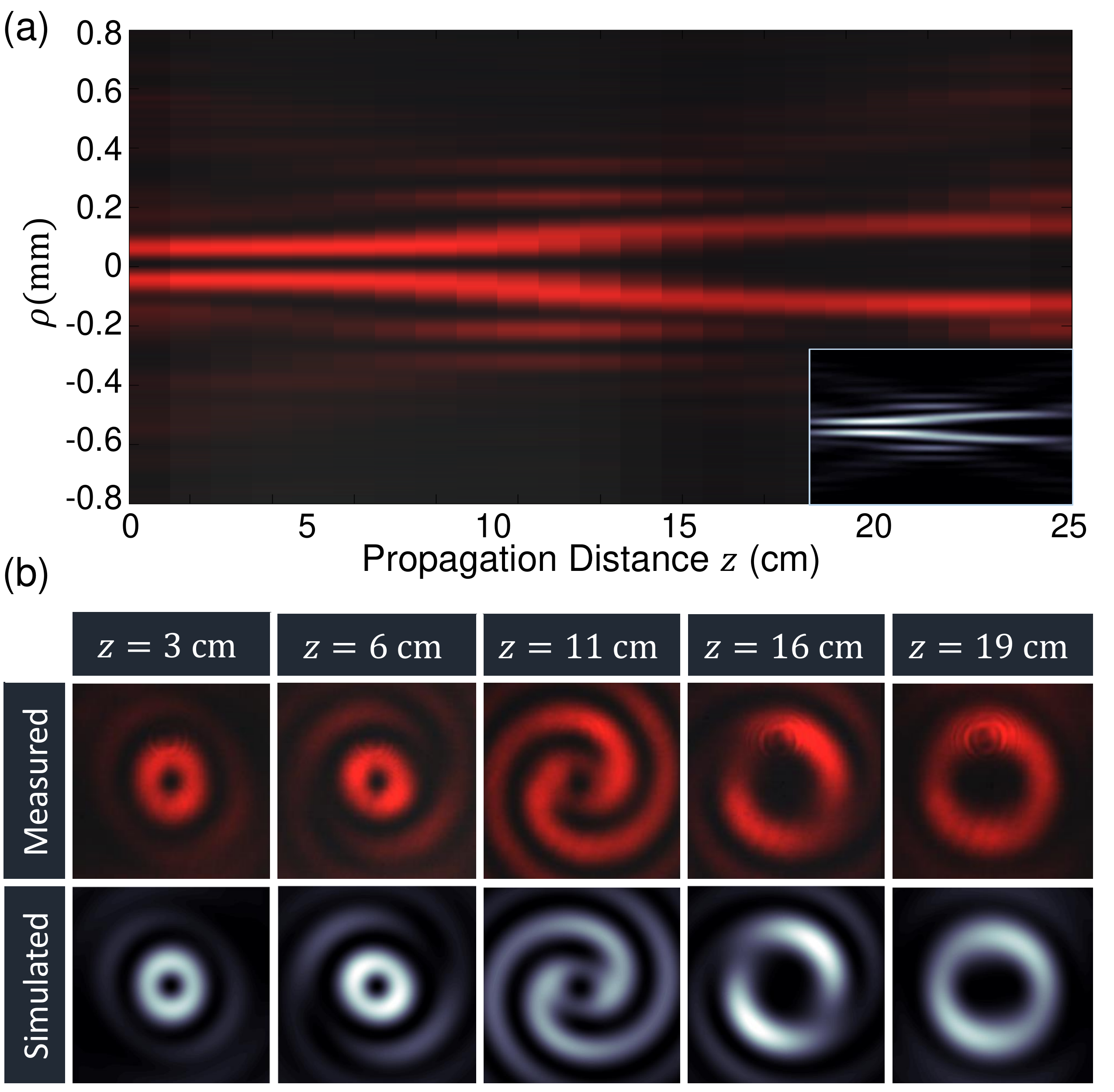}
   
\caption{Measured and simulated intensity of the beam varying its topological charge. a) Measured longitudinal intensity profile (the inset depicts the simulated intensity profile). b) Measured and simulated transverse intensity profiles obtained at propagation distances: $z=3$ cm, $z=6$ cm, $z=11$ cm, $z=16$ cm, and $z=19$ cm.}
	\label{Fig7}
\end{figure}

Note that we assigned slightly larger value for $F_{3}$ over the interval (11 cm  $\leq z \leq$ 23 cm) to ensure that $\psi_1$ and $\psi_3$ are generated at equal intensity levels, i.e. to compensate for the energy mismatch between their Bessel modes. 

Based on Eq.~(\ref{Eq13}), the beam is designed to have a topological charge $\ell=1$ over the interval (0 cm  $\leq z \leq$ 11 cm), and then experiences a transition in its topological charge at $z=11$ cm, from $\ell=1$ to $\ell=3$. The beam then maintains this charge ($\ell=3$) for the remaining distance (11 cm  $\leq z \leq$ 23 cm) before it switches off. Figure~\ref{Fig7} shows the intensity profiles of the generated beam. The longitudinal intensity profile has been rendered by aggregating 1D slices obtained from the transverse CCD images taken over 30 cm (with 1 cm steps) in the $z$ direction, and is in good agreement with the simulated results shown in the inset. The transverse profiles are shown in Fig.~\ref{Fig8}(b) at different planes along the direction of propagation. Note how the vortex beam diameter is increased at $z=19$ cm --- a signature of increasing the topological charge. In addition, similar to the case of topological inversion, here, there exist a region in space (at $z=11$ cm) where the closed ring of the vortex undergoes shape deformation. This deformation is a characteristic of topological transition. It can be attributed to the spatial redistribution incurred to the phase singularities, associated with varying the topological charge. In essence, the intensity profile is deformed judiciously to allow certain number of charges to enter to (or exit from) the beam's center to satisfy Eq.~(\ref{Eq13}), as will be shown.

\subsection{Mechanism of topological charge transition}

\begin{figure*}
	\centering
	\includegraphics[width=1\textwidth]{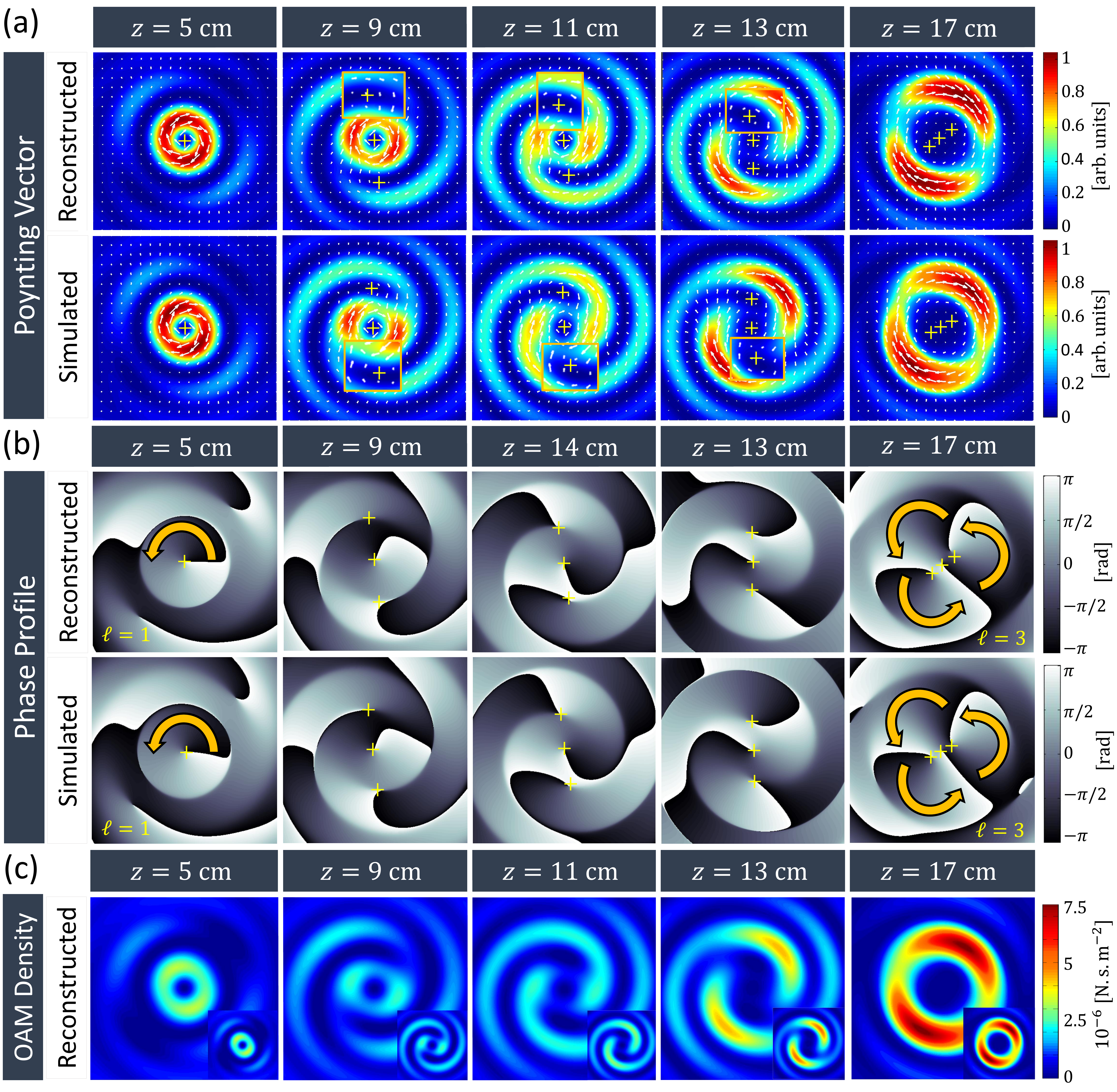}
    \caption{Theoretical and experimental evolution of the Poynting vector, phase, and OAM density of our 3D structured light varying its charge.  All plots reveal the transverse plane at various propagation distances, $z$. a) Transverse components of the Poynting vector. The white arrows trace the direction of energy flow and the `+' and `-' signs denote the polarity of the local vortices. The boxed areas represent enlarged sections of the image. b) Reconstructed and simulated phase profile. The arrows depict the sense of helicity in the phase and their number denote the topological charge. c) OAM density profiles reconstructed at different propagation distances. The insets depict the simulated profiles. The OAM density triples its magnitude as the beam propagates.}
  	\label{Fig8}
\end{figure*}

To investigate the mechanism by which the vortex beam undergoes topological charge transition, we obtained the Poynting vector of the reconstructed field. More details on the modal decomposition and reconstruction of this beam at various planes along the propagation direction, $z$, can be found in Appendix B. Figure~\ref{Fig8}(a) shows the transverse components of the Poynting vector of the reconstructed beam at multiple planes along the beam's axis. At the plane $z=5$ cm, the energy circulates in the clock-wise direction, over a small ring, where the charge is supposedly $\ell=1$. Eventually, at $z=17$ cm, the charge has evolved from $\ell=1$ to $\ell=3$, signified by the energy circulation over a larger ring. Such topological transition is also confirmed by looking at the reconstructed phase profiles depicted in Fig.~\ref{Fig8}(b); from which it is clear that the number of helical phase-fronts is increased from 1 to 3, while maintaining the same sense of helicity, as the beam propagates. 

Evolution of the topological charge from $\ell=1$ to $\ell=3$ relies on several interesting dynamics that are reminiscent of those associated with topological inversion. First, the closed ring forming the vortex beam (with $\ell=1$) gradually splits into two sections. This deformation is accompanied by the formation of two smaller vortices around the beam's center, signified by the `+' sign in Fig.~\ref{Fig8}(a). It is also associated with the creation of two new singularities in the phase-front, which gradually acquire an azimuthal phase gradient as the beam propagates. Second, the newly formed vortices (charges) approach the beam's center and come into the vicinity of the original vortex. Third, as the three vortices (charges) come closer, the divided ring gradually merges again, but this time into a larger diameter, consistent with increasing its charge. Simultaneously, the phase-front evolves into three intertwined helices --- a signature of acquiring a charge $\ell=3$. Interestingly, the intertwined helices possess three distinguished singularities connected via branch cuts as opposed to sharing one singularity. 

Shape deformation observed here is a generic behavior and can be regarded as a signature of topological transition. In any structured (non-trivial) transition from $\ell_1$ to $\ell_2$, the beam undergoes shape deformation, thus creating $|\ell_2-\ell_1|$ channels of zero intensity allowing to transport charges to (from) the beam's center. This observation holds true in the case of topological inversion as well. These results advance prior work on single component beams vortex dynamics by Gouy phase \cite{baumann2009propagation}, confirming the prediction by the authors of the anticipated rich dynamics of superposition fields.

The OAM density of the reconstructed beam has been obtained via Eq.~(\ref{Eq10}), and is depicted in Fig.~\ref{Fig8}(c) at three planes along the $z$ direction. Note how the OAM density acquires larger values, three times its initial value, and gets redistributed over larger diameter as the charge evolves from $\ell=1$ to $\ell=3$. A discussion on OAM and charge conservation is presented next. 

\subsection{Conservation of OAM and charge}

The transition from $\ell=1$ to $\ell=3$ in our 3D structured beam occurs as a result of controlled creation and movement of two phase singularities in the vicinity of the beam's center as it propagates. 
\begin{figure}[ht!]
	\centering
	\includegraphics[width=0.48\textwidth]{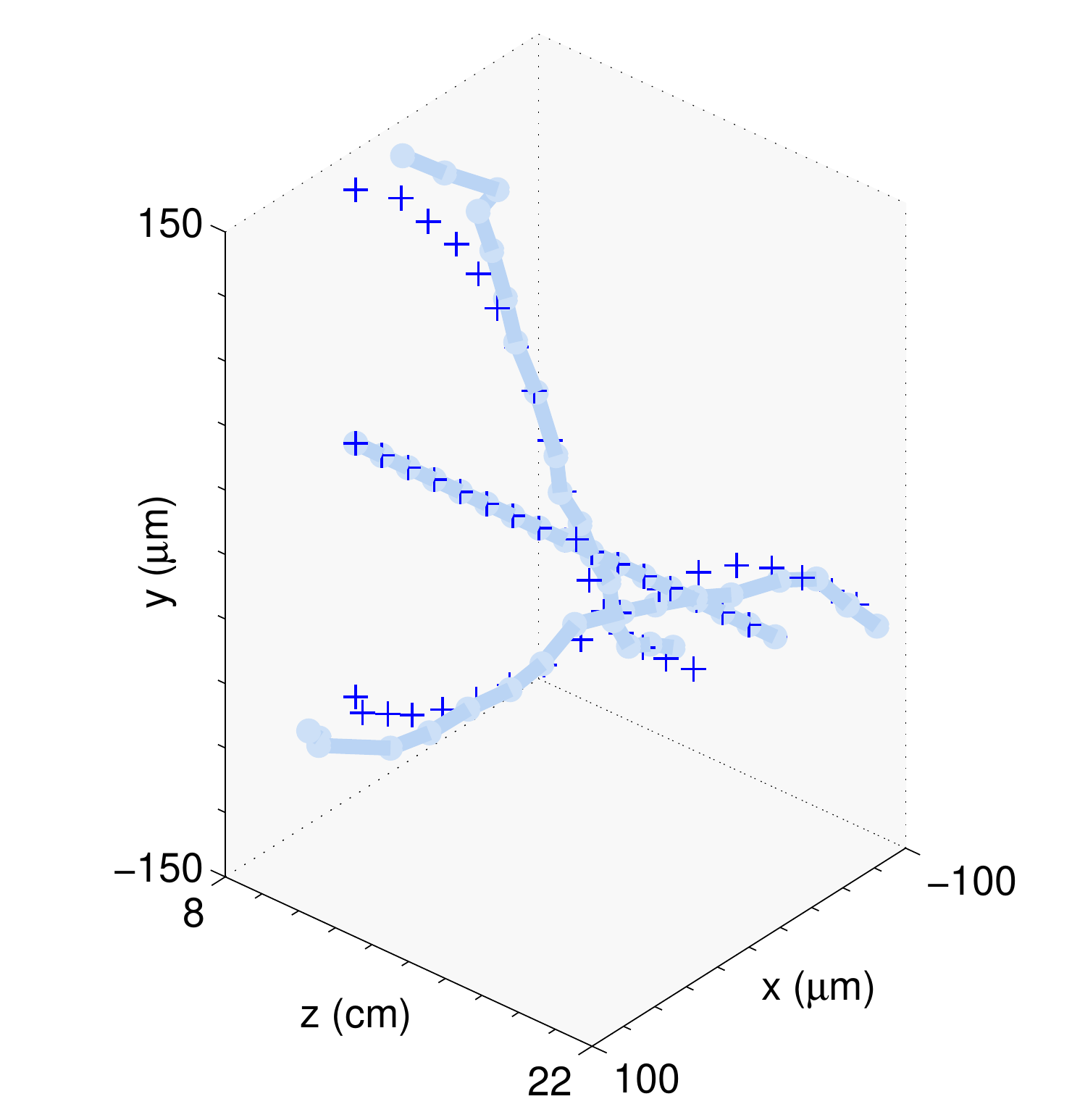}
\caption{Movement of charges in the vicinity of the beam's center. The solid lines depict the experimentally reconstructed paths whereas the markers represent the theoretical paths.}
	\label{Fig9}
\end{figure}

Figure~\ref{Fig9} illustrates the 3D path traversed by the charges near the beam's center as it undergoes the topological transition. The charges created in the outer rings of the beam are transported into the beam's center at the prescribed distance, $z=11$ cm, to satisfy the desired OAM profile dictated by Eq.~\ref{Eq13}.

To examine OAM conservation, we evaluated the OAM of the beam, given by Eq.~(\ref{Eq11}) over two aperture diameters: a) $D=400$ $\mu$m, and b) $D=5$ mm. Additionally, we computed the effective topological charge $\ell_{\mbox{eff}}$ from Eq.~(\ref{Eq12}) considering the same two aperture sizes.

\begin{figure}[ht!]
	\centering
	\includegraphics[width=0.42\textwidth]{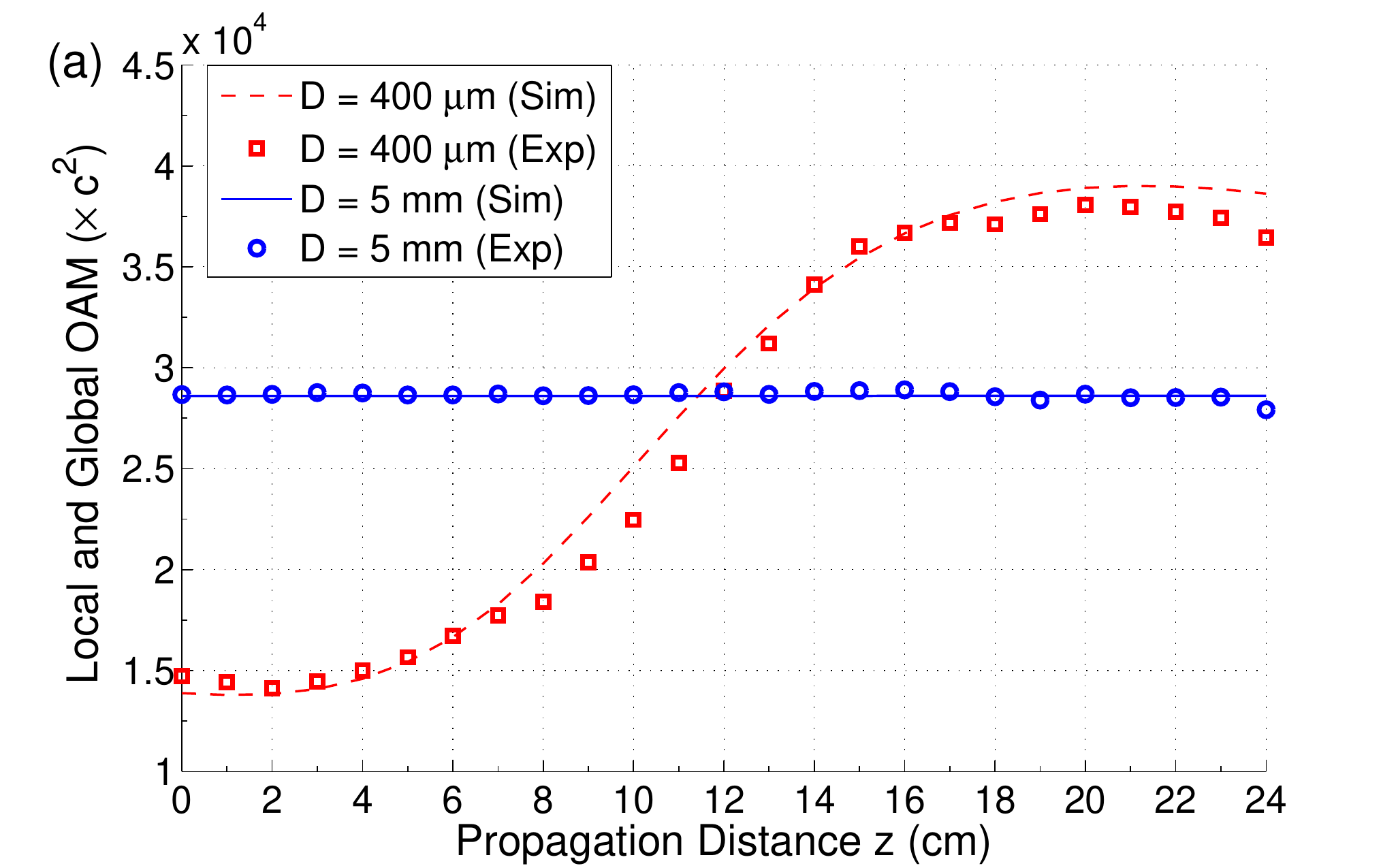}
	\includegraphics[width=0.42\textwidth]{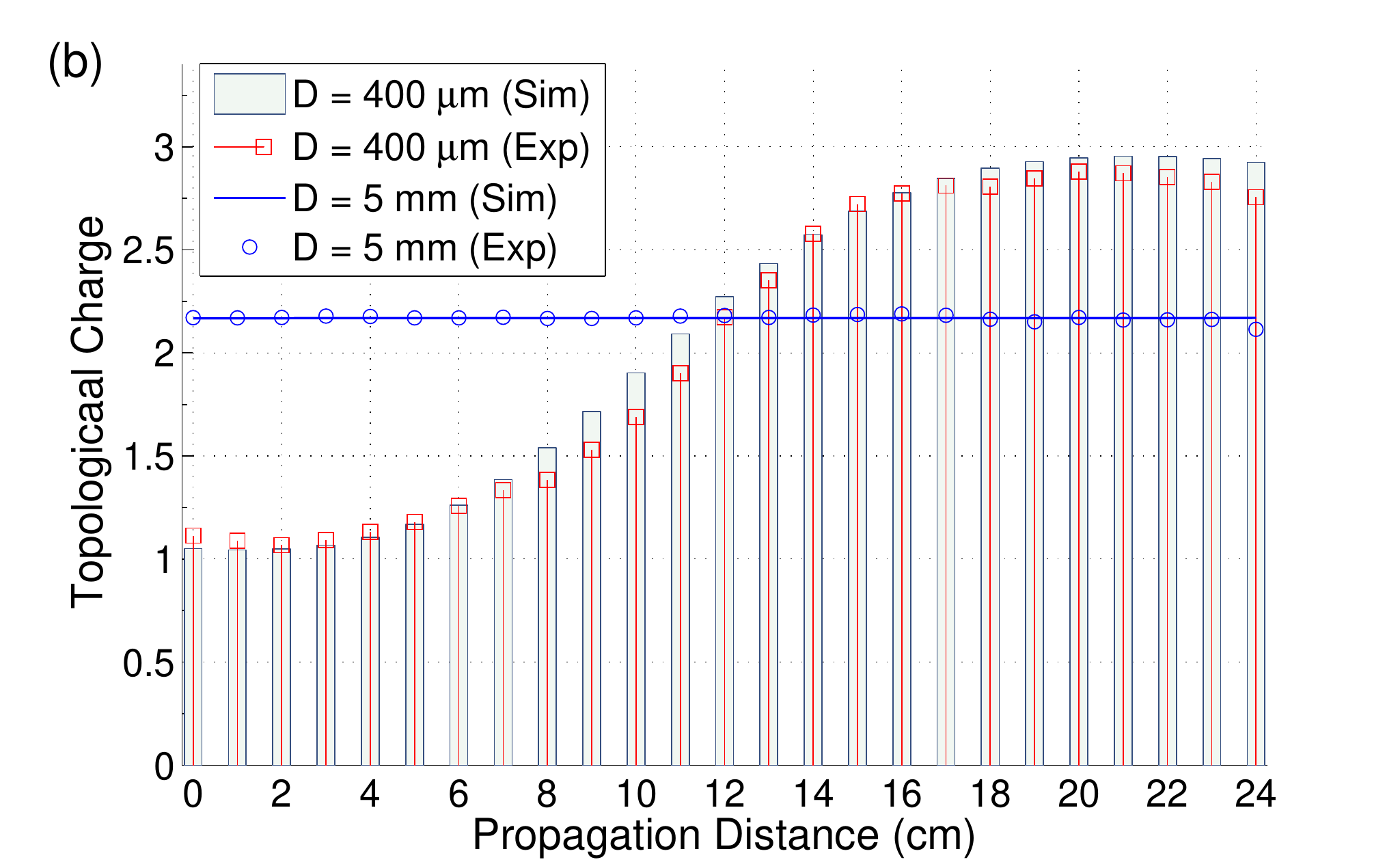}
\caption{a) Evolution of the OAM and topological charge considering two aperture diameters $D=400$ $\mu$m, and $D=5$ mm. a) Local and global OAM of the beam obtained along its direction of propagation. b) Local and global effective topological charge evaluated along the beam's axis.}
	\label{Fig10}
\end{figure}

The OAM per unit length ($J_z$) and $\ell_{\mbox{eff}}$ are depicted in Fig~\ref{Fig10}(a) and (b), respectively. The experimental results are compared with the simulations and are in very good agreement. In the limit of small aperture size (excluding the outer rings of the beam), the OAM exhibits a transition into larger values as it propagates. Therefore, the OAM conservation is seemingly broken \textit{locally}, i.e. over this finite section of the beam. However, when the OAM is evaluated across the entire cross section of the beam ($D=5$ mm), it maintains a fixed value regardless of the propagation distance $z$. The \textit{global} OAM maintains a value that lies in between the minimum and maximum local OAM values, and is always conserved in this case. Similar to the case of topological inversion, here, while the OAM varies locally, the global OAM is always conserved. A similar picture is also observed in the effective topological charge $\ell_{\mbox{eff}}$, as shown in Figure~\ref{Fig10}(b). While the local charge evolves from $\ell_{\mbox{eff}}=1$ to approach $\ell_{\mbox{eff}}=3$, the global $\ell_{\mbox{eff}}$ remains fixed at $\ell_{\mbox{eff}}=2.15$ at all planes along the $z$ direction. Note that $\ell_{\mbox{eff}}$ is slightly larger than 2 (the average of $\ell=1$ and $\ell=3$) due to the asymmetric weightings assigned to $\psi_1$ and $\psi_3$ in Eq.~(\ref{Eq13}).  

These interpretations are general and prove that while the total number of charges entering/exiting an open surface within the beam can vary, the total number of charges carried by the entire system is always in equilibrium, provided that it is a closed system. In this case, local topological transitions can occur as a result of spatial redistribution of the charges and the local OAM densities without altering the respective global quantities. In fact, while the OAM density ($j_z$) and the total OAM per unit length ($J_z$) can vary along the beam's axis when observed over a finite open aperture, it is worth noting that $j_z$ together with its associated flux will always satisfy the continuity equation when observed over a closed surface (enclosing a volume). 

\section{Discussion}
Conservation of OAM and charge manifests in a manner that can be regarded as a practical realization of the Hilbert's Hotel paradox \cite{potovcek2015quantum,Gbur:16, wang2017hilbert}. The paradox, attributed to David Hilbert in 1924, postulates a hotel with infinitely countable number of occupied rooms with no vacancies. Nevertheless, such a hotel can still accommodate an infinite number of new guests at any given time by merely shifting each guest to its neighbor higher-numbered room, thus creating a seemingly infinite number of vacancies. Similarly, here, the accommodation of new positive charges (new guests) in the beam's center is associated with the formation of negative charges (vacancies) in the outer rings of the beam (and vice versa). In this case, the outer rings act as a reservoir that dynamically compensates for the controlled imbalance in both the charges and OAM within the beam's center. The underlying mechanism of charge conservation thus exhibits a fascinating analogy with the seemingly abstract concept of Hilbert's hotel.  

Furthermore, this work presents a systematic approach --- expressed in closed-form analytic expressions --- to design, create, and characterize the 3D structured beams using a simple all-optical holographic setup. This significantly expands some earlier efforts that relied on complicated setups incorporating non-linear media to achieve non-trivial topological inversion.  Moreover, we have employed modal decomposition in pseudo-orthonormal function space to realise a holistic quantitative diagnostic tool for our structured light, determining the intensity, phase, wavefront, Poynting vector and OAM density.  As such we are able to offer a complete interpretation of the OAM dynamics in such 3D structured light beams.  In particular, we established that at any structured transition from $\ell_1$ to  $\ell_2$, the beam undergoes deliberate shape deformation to create exactly $|\ell_2 - \ell_1|$ channels of zero intensity allowing to judiciously transport charges to (from) the beam's center. This is a generic signature that we emphasize here for the first time and is a step towards establishing other governing laws in advanced singular optics.

Additionally, we point out that while here we have demonstrated cases where either the sign or the magnitude of $\ell$ is changed with propagation, simultaneous control of both is readily achievable as well. Further, in this work, we focused on the orbital component of the angular momentum by considering linearly polarized light. Indeed, the photon's spin, associated with circular polarization, also contributes to the total angular momentum \cite{ALLEN200067}. Extensive research efforts have been dedicated to the spin-orbit coupling and its role in creating vortex modes \cite{PhysRevLett.96.163905}; it will be interesting to investigate the dynamics associated with spin-orbit coupling using the 3D structured light beams demonstrated herein. 

\vspace{0.5cm}

\section{Conclusion}
We have created and studied three dimensional structured light that undergoes arbitrary (but controllable) transitions in their topological charges and OAM along their axis of propagation. This class of beams has been realized via $z$-dependent interference of multiple Bessel modes, demonstrating counter-intuitive cases of OAM evolution in free-space. Using all-optical modal decomposition, we reconstructed the structured beam and obtained its Poynting vector, phase, and OAM density, thus revealing the underlying physical mechanisms governing the beam's evolution. In particular, we were able to monitor the movement, creation, and annihilation of optical vortices, which accounts for the local variation of the OAM density within regions of our 3D field. Our results show that while OAM is conserved globally, it may be arbitrarily tailored locally, even violating OAM conservation within a finite section of the beam. This study provides new insights into OAM in structured light fields that may be useful in understanding topological deformations, opening new possibilities in dense data communications, micro-manipulation, and remote sensing.

\section*{Acknowledgment}	
The authors wish to thank Bienvenu Ndagano and Isaac Nape for their stimulating discussions.  We also acknowledge the support from the Natural Sciences and Engineering Research Council of Canada (NSERC) and the Ontario Graduate Scholarship (OGS) program.

\clearpage
\section*{Appendices}	
\begin{appendix}
Appendices A and B contain additional data that are useful for the proper implementation of modal decomposition and reconstruction of optical fields using the Bessel modes. Appendix A presents cross-talk analysis to assess the orthogonality and overlap between the Bessel functions, whereas Appendix B includes the results of modal decomposition for the second 3D structured beam discussed in the main article --- changing its topological charge form $\ell=1$ to $\ell=3$.
\section{Cross-Talk Measurements}
A necessary requirement for performing the all-optical modal decomposition of structured beams is to adopt orthonormal basis functions \cite{Kaiser:09,Flamm:12,Dudley:14,1367-2630-15-7-073025,Flamm2013-2,rosales2017shape}. In essence, orthogonality defines the degree of \textit{distinguish-ability} of each mode (basis function) at the detection system, whereas normalization ensures that the total energy of the optical field is conserved when transformed from one basis system to another. The Bessel functions deployed in our 3D structured beams, however, are in general not orthogonal in nature. Nevertheless, one can still obtain a normalized \textit{pseudo-orthogonal} set of Bessel functions provided that their transverse wavenumbers $k_{\rho}$ are sufficiently separated, as will be discussed shortly. 

To assess the degree of orthogonality within the Bessel functions, we performed multiple cross-talk analysis. Here, cross-talk represents the amount of overlap between one mode and another, and is mathematically represented by 
\begin{equation}
\int\int_{\mathbb{R}^2} V^*_{j}\cdot V_{j'} \rho d\rho d\phi.
\end{equation}

\begin{figure}[!htb]
	\centering
	\includegraphics[width=0.375\textwidth]{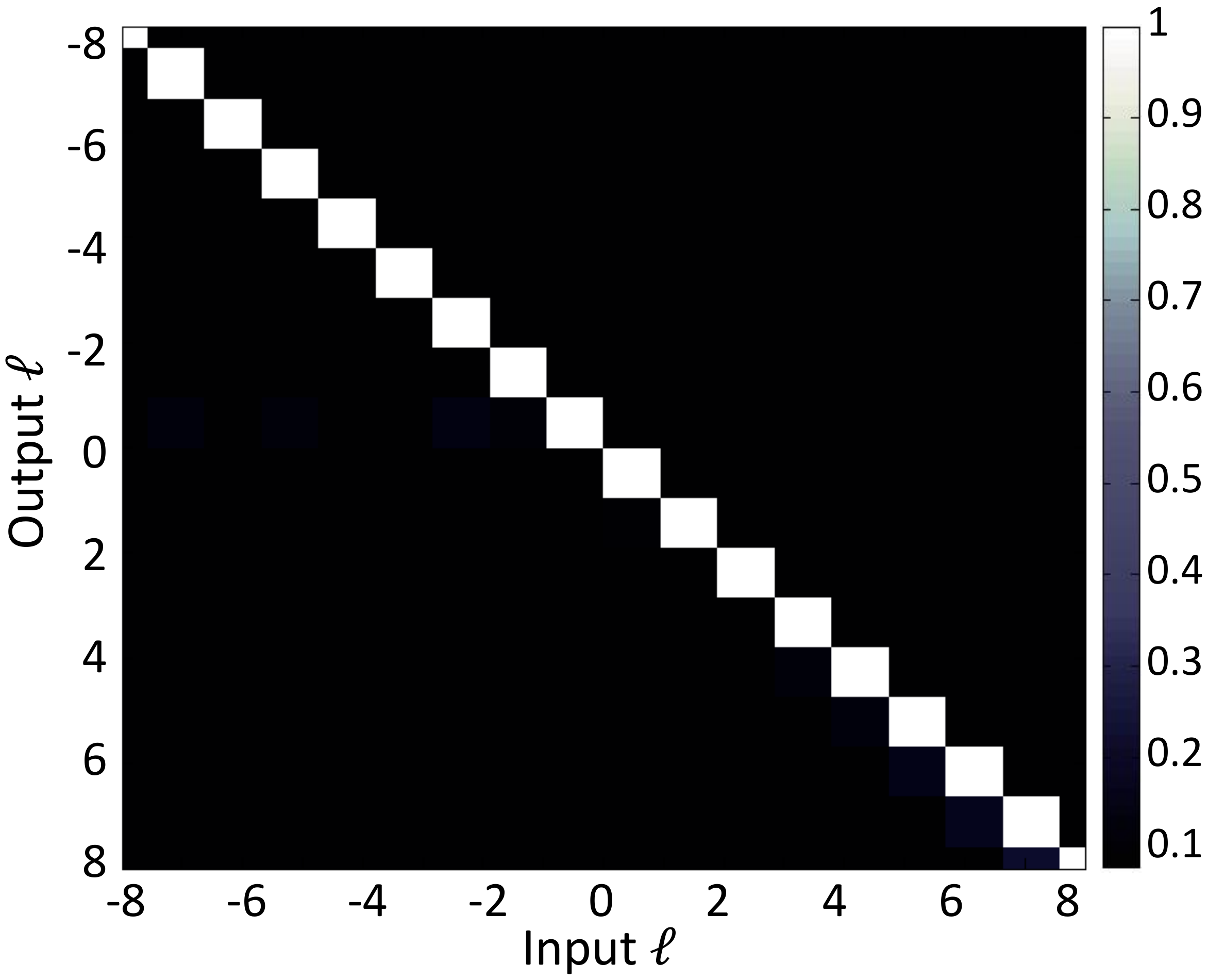}
	\caption{Cross-talk measurement for Bessel beams with different values of $\ell$, while fixing the value of $k_{\rho}$ at $k_{\rho}=3.48\times10^4$ m$^{-1}$.}
	\label{FigS1}
\end{figure}

Ideally, when the modes in the function space are orthogonal, the cross-talk should be zero as long as $j \neq j'$. In our case, each Bessel function $J_{\ell}(k_{\rho})e^{i\ell\phi}$ is defined by its radial dependence $k_{\rho}$ and azimuthal dependence $e^{i\ell\phi}$. To create a discrete set of pseudo-orthogonal Bessel functions, we start by analyzing the cross-talk among the Bessel modes with different values of $\ell$, while fixing the parameter $k_{\rho}$. Figure~\ref{FigS1} depicts the measured cross-talk between the Bessel modes with different quantized charges $\ell$. The maximum cross-talk is less than -10 dB in this case, which implies that the adopted Bessel modes are sufficiently distinguishable in the azimuthal direction.

In addition to the radial dependence, the vortex Bessel modes are also characterized by a radial dependency described by the continuous variable $k_{\rho}$. We introduce the metric $\delta k_{\rho}$ to describe the separation between Bessel functions with different $k_{\rho}$, in percentage form, such that 
\begin{equation}
\delta k_{\rho} = \frac{k_{\rho}^{j_1} -k_{\rho}^{j_2} }{k_{\rho}^{j_1}}\times 100. 
\end{equation}

\begin{figure*}[!htb]
	\centering
	\includegraphics[width=0.725\textwidth]{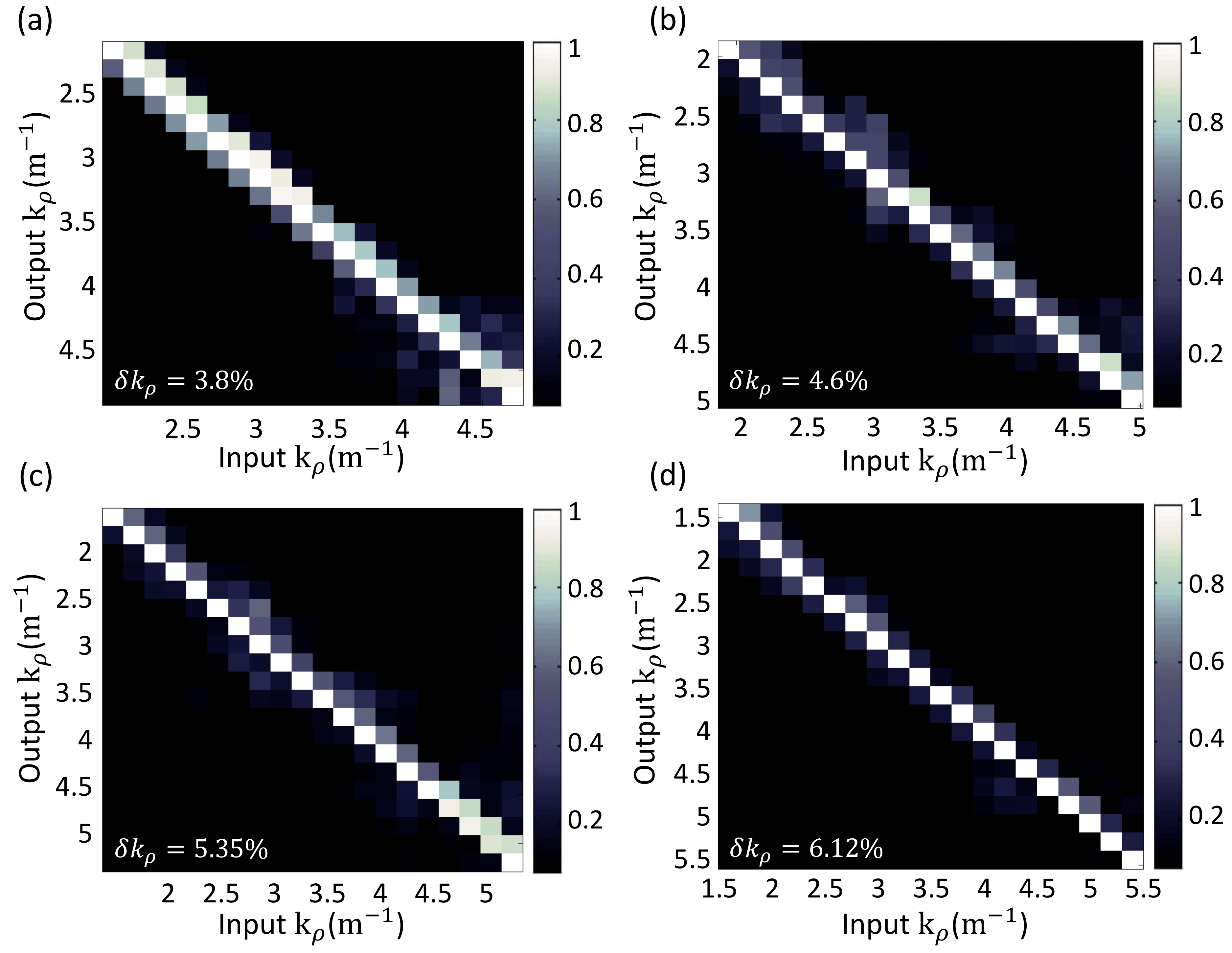}
	\includegraphics[width=0.725\textwidth]{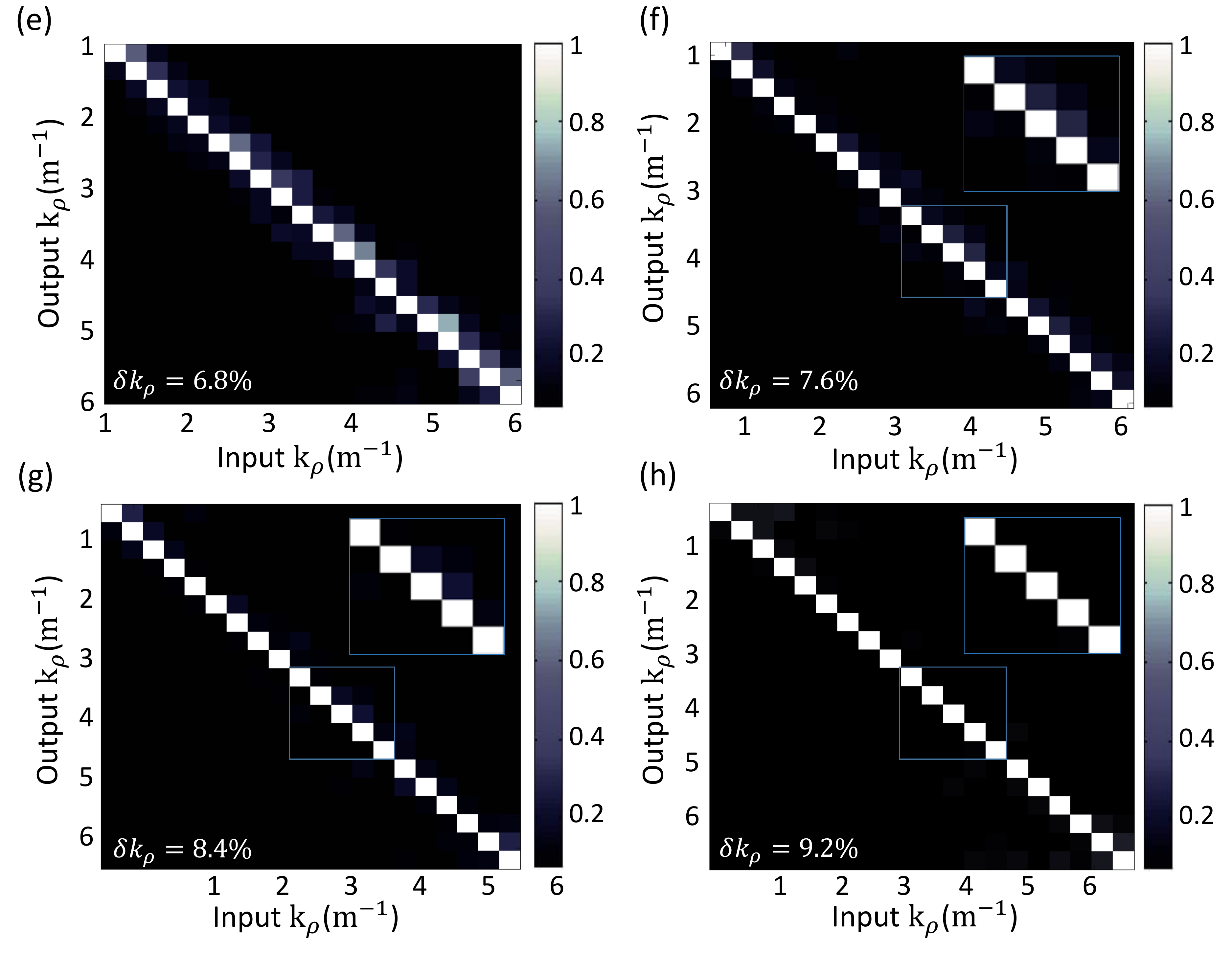}
	\caption{Cross-talk measurements for Bessel beams as function of $\delta k_{\rho}$: a) $\delta k_{\rho}=3.8\%$, b) $\delta k_{\rho}=4.6\%$, c) $\delta k_{\rho}=5.35\%$, d) $\delta k_{\rho}=6.12\%$, e) $\delta k_{\rho}=6.8\%$, f) $\delta k_{\rho}=7.6\%$, g) $\delta k_{\rho}=8.4\%$, h) $\delta k_{\rho}=9.2\%$. }
	\label{FigS2}
\end{figure*}

This defines the distance between one Bessel function and another in the k-space. Figure~\ref{FigS2} depicts the cross-talk measurements under different separations of $k_{\rho}$. 
It is observed that, while the cross-talk is remarkable in case (a), it is significantly reduced as the separation $\delta k_{\rho}$ is increased. For instance, at $\delta k_{\rho}=9.2\%$, the cross-talk is less than -10 dB which implies that the Bessel functions are distinguishable in the radial direction in that case. Accordingly, in all the 3D structured beam profiles presented in the main article, we ensured that the separation among the Bessel functions in the k-space always satisfies the condition $\delta k_{\rho} \geq 9.2\%$. 

Consequently, it is possible to obtain a discrete set of Bessel functions that are pseudo-orthogonal in both the radial and azimuthal directions. Once this set is defined, each Bessel mode can then be normalized with respect to its total energy to satisfy the orthonormality condition and perform the modal decomposition efficiently.

\section{Spectral decomposition for the second generated 3D pattern}

To access key quantities of the second generated beam such as its Poynting vector and OAM, we performed full modal decomposition and reconstruction of the field into its Bessel basis along the propagation direction. Modal decomposition  also provides useful insights into the dynamics of topological transition of the 3D structured beam from $\ell=1$ to $\ell=3$. 

\begin{figure}[!htb]
	\centering
	\includegraphics[width=0.48\textwidth]{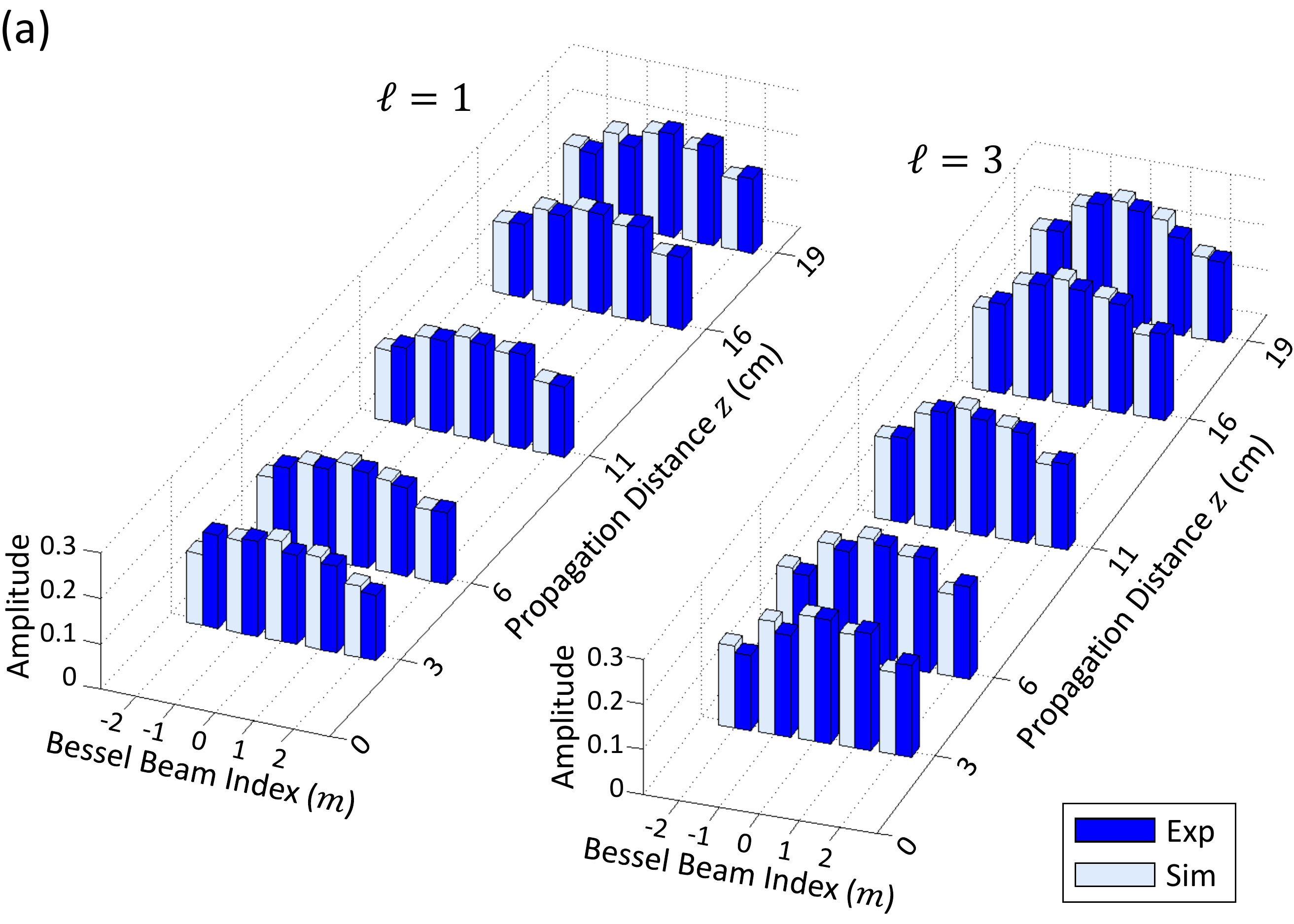}
	\includegraphics[width=0.48\textwidth]{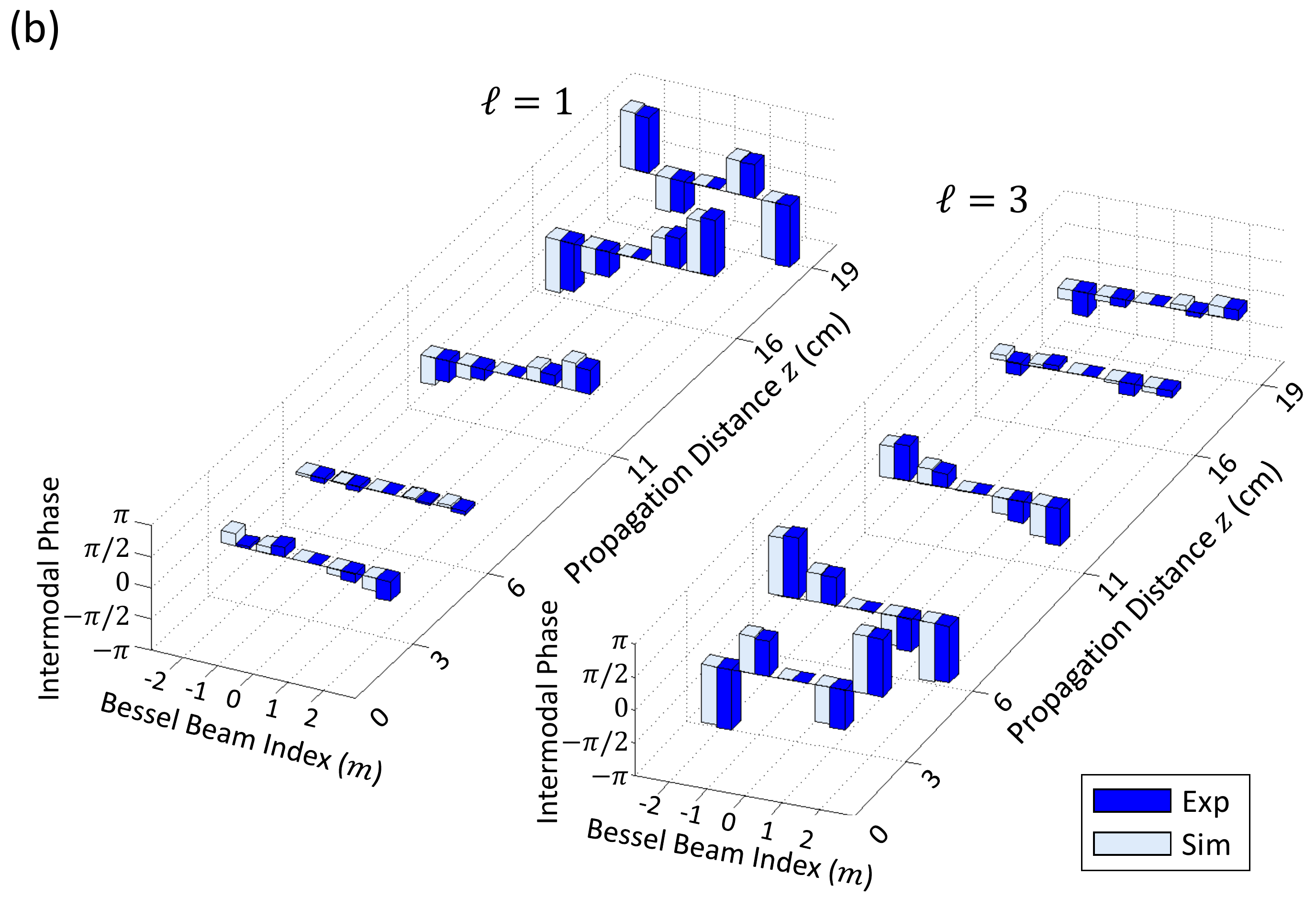}
	\caption{Spectral decomposition of the generated 3D structured light into Bessel basis. a) Amplitudes of $A_{1, m}$, and $A_{3, m}$, b) intermodal phases among the Bessel modes within $\psi_1$ and $\psi_{3}$. The coefficients are displayed at $z=3$ cm, $z=6$ cm, $z=11$ cm, $z=16$ cm, and $z=19$ cm.}
	\label{FigS3}
\end{figure}

Figure~\ref{FigS3} depicts the measured amplitudes and intermodal phases among the Bessel beams of $U(\rho,\phi,z,t)$ in comparison with the theoretical predictions. The modal coefficients are displayed at five distances: $z=3$ cm, $z=6$ cm,, $z=11$ cm, $z=16$ cm, and $z=19$ cm, selected as illustrative samples for the 3D structured beam's evolution. 

Evidently, the amplitudes $|A_{\ell, m}|$ remain constant throughout the beam propagation, whereas the intermodal phases (proportional to $\angle A_{\ell, m}e^{i k_z^{\ell, m}z})$ vary with propagation, thus shaping the longitudinal beam profile. The frame at $z=3$ cm in Fig.~\ref{FigS3} captures the intermodal phases within $\psi_1$ as it progresses to become in-phase later on at $z=6$ cm. The Bessel beams of $\psi_3$, on the other hand, are out-of-phase at this position. They destructively interfere and their energies are dispersed into the outer rings of the beam. Hence, they do not contribute to the beam's center and the topological charge thereby becomes predominately $\ell=1$. The opposite picture is seen later on at $z=16$ cm and $z=19$ cm; where the Bessel beams associated with $\psi_{3}$ progress to become in-phase and those associated with $\psi_1$ become out-of-phase. Hence, the topological charge $\ell$ evolves from $\ell=1$ to $\ell=3$. Certainly, there is an overlap region, which exists at $z=11$ cm, and where the contributions of both $\psi_1$ and $\psi_{3}$ are present in the beam's center, and where the topological transition takes place.

\end{appendix}

\bibliography{bibfile}

\end{document}